 \documentclass[final,5p,times,twocolumn,authoryear]{elsarticle}

\usepackage{amssymb}
\usepackage{lipsum}
\usepackage{url}
\usepackage{xcolor}
\journal{Astronomy $\&$ Computing}

\begin{document}

\begin{frontmatter}



\title{Analysis of Chiral Oxirane Molecules in preparation for Next Generation Telescopes: A Review, New Analysis, \& a Chiral Molecule Database}


\author[1,2]{C.D. Tremblay}
\affiliation[1]{SETI Institute, 339 Bernardo Ave, Suite 200, Mountain View, CA 94043, USA}
\affiliation[2]{Berkeley SETI Research Center, University of California, Berkeley, CA 94720, USA}

\author[3]{R.D. Amos}
\affiliation[3]{University of New South Wales, Canberra, Australia}

\author[4]{R. Kobayashi}
\affiliation[4]{Australian National University, Canberra, Australia}

\begin{abstract}
Human biology has a preference for left-handed chiral molecules and an outstanding question is if this is imposed through astrophysical origins. We aim to evaluate the known information about chiral molecules within astrophysical and astrochemical databases, evaluate chemical modeling accuracy, and use high-level CCSD(T) calculations to characterize propylene oxide and other oxirane variants. By comparing these computational values with past laboratory experiments, we find a 99.9\% similarity.  We also have put together a new database dedicated to chiral molecules and variants of chiral molecules to assist in answering this question.
\end{abstract}




\end{frontmatter}

%\tableofcontents



\section{Introduction}
\label{introduction}

Life on Earth is homochiral, preferring one enantiomer of chiral molecules \citep{cshlpOriginBiological,CM20}. It is unknown whether pre-biotic chemistry required a pre-existing excess of one chiral form over the other or if it created the excess \citep{GJ10}. Such an excess has been measured in meteorites \citep{Myr15,Coop21} hinting at an extra-terrestrial origin of homochirality. Despite the catalog of interstellar molecules growing to more than 200 entries and several dedicated searches, only one chiral molecule, propylene oxide\footnote{Carbodiimide (HN=C=NH), an isomer of cyanamide was detected by \cite{Turner_75} and in principle is chiral. However, in practice it likely converts between the two forms rapidly enough that the chirality is not observable and therefore is not considered here.}, has been detected outside our solar system \citep{McG16} and the identification of any other chiral molecule is an ongoing challenge \citep{Ell20}. With the soon-to-be-constructed Square Kilometre Array (SKA) in the southern hemisphere and the next-generation Very Large Array (ng-VLA) in the northern hemisphere, along with current operational precursor and pathfinder telescopes, we now have the tools to discover our potential cosmic chiral origins.

Over the last several years many of the SKA pathfinder and precursor telescopes, including the Murchison Widefield Array \citep{Way18,Mor23} and the Karl G. Jansky Very Large Array \citep{evla_11}, completed significant upgrades with some focus on increased capabilities in astrochemistry. This includes larger bandwidth, better frequency resolution, and increased sensitivities and has permitted several new discoveries ({\it e.g.} \cite{Tre18}, \cite{De21}). Additionally, these upgrades allowed for a better understanding of the computational requirements needed for large spectral and regional blind molecular surveys using upcoming, next-generation telescopes.

Currently, the SKA Low-Frequency Array in Australia (50--500\,MHz) and the SKA Mid-Frequency Array in South Africa (500--15000\,MHz) are under Phase I construction. The ng-VLA in the United States (1.2--116\,GHz) is under development, with construction anticipated to start in 2026. All three of these instruments have key science goals of understanding the origins, prevalence, and evolution of life in the Universe ({\it e.g.}\citep{SKACOL,McG18})\footnote{\url{https://www.skao.int/en/explore/science-goals/145/seeking-origins-life}\\ 
\url{https://public.nrao.edu/gallery/cosmic-origins-of-life-ngvla-key-science-goal-2/}}. 

Therefore this is an opportune time to build and compile expected rotational transitions that radio telescopes would be sensitive to that these future telescopes could discover.  With the emergence of significantly upgraded telescopes in infrared, like the $James$ $Webb$ $Space$ $Telescope $ (JWST), knowledge of our collective understanding of vibrational transitions is also necessary.

Computational chemistry has a strong track record in aiding astrochemists and astronomers by simulating experimental data with sufficient precision to identify molecules found in interstellar regions. The detection of such molecules relies on high-resolution rotational spectra, which correspond to the wavelength region of the radio telescope surveys. Although rapid methods such as those explained in \cite{Zapata_2021} or \cite{Lee2020} provide insight into which proposals for relevant frequencies should be covered by new facilities, they do not provide the detailed information required to start an observational campaign. Current large blind surveys for complex organic molecules like ``A Rigorous K/Ka-Band Survey Hunting for Aromatic Molecule" (ARKAM) and ``Green Bank Telescope (GBT) Observations of TMC-1: Hunting Aromatic Molecules" (GOTHAM;\citealt{GOTHAM}) are already finding large reservoirs of gas around star forming regions of the interstellar medium (ISM) but more precise models will be required for the future generation of telescopes.

Various groups have put a focus on providing detailed computational models for complex organic molecules\footnote{Carbon-containing molecules with 6 atoms or more.}, especially those that may indicate life (biosignatures). This has included not only an accurate understanding of the transition frequencies and chemical abundance but also the pathways in which the chemicals may form ({\it e.g.} \cite{Coutens2022, Dze22}). For example, \cite{Paulive2021} investigated C$_2$H$_4$O$_2$ isomers and their potential relationship to cold dark molecular environments through radiolysis (a process of molecular damage to a substance caused by ionizing radiation). They found that by adding in radiolysis, the abundances of selected molecules would be greatly increased.

More recently, \cite{Alessandrini2023} created a computational approach to discover the most stable forms of C$_3$H$_3$NO isomers. This molecular structure is theorized as being important to prebiotic chemistry but has yet to be discovered in the interstellar medium \citep{Dickens1996}.  \cite{Alessandrini2023} approach uses a double-hybrid DFT (density functional theory) in conjunction with a partially augmented triple-zeta basis set to evaluate and prioritize transitions to focus astronomical observations on and the frequencies that are potentially detectable.

With this in mind, we have undertaken to investigate computationally chiral molecules that are potentially present in interstellar space. There exist already many databases containing information on interstellar molecules as described in the next section. From these databases and a search of the general literature, we were able to find several candidates worthy of further investigation.

\begin{enumerate}

\item[i.] Propylene oxide\\

In \cite{McG16} they reported the astronomical detection of propylene oxide in the gas phase around protostellar clusters in the Sagittarius B2 star-forming region. Experimental rotational constants were already determined in 1977 by \cite{CS77}, and updated by \cite{Mesko}. Computationally, \cite{Ell20} calculated structural and spectroscopic properties of propylene oxide and variants using density functional theory (DFT) in the quest for finding more chiral molecules in the interstellar medium. The most detailed theoretical study of the vibrational spectrum is that by \cite{BBBP14}.

\item[ii.] 2-aminopropionitrile\\

\cite{Mol12} proposed 2-aminopropionitrile as a good candidate for interstellar detection based on the detection of aminoacetonitrile in Sagittarius B2 by \cite{Bel08}. Accordingly, they provide experimental microwave and sub-millimeter spectra (8–-80\,GHz, 150–-660\,GHz) and high-level {\it ab initio} calculations for the most stable conformer and its five lowest excited vibrational states.

\item [iii.] Tartaric acid \\

Tartaric acid is the iconic molecule with which \cite{Pasteur} established the existence of chirality and is likely to exist in the interstellar medium as it has already been found in the Murchison meteorite \citep{Coop21}.

\item[iv.] alpha-aminoethanol \\

\cite{Duv10} proposed alpha-aminoethanol, an amino acid precursor, as a possible link between interstellar molecules and life on Earth.

\item [v.] C2H5NO isomers\\

A recent paper by \cite{Simmie22} investigated isomers of acetamide, one of the most abundant of the molecules present around Sagittarius B2. Of the various isomers considered, several are chiral: 2-aziridinol (or 2-hydroxyaziridine) and 2-methyl-oxaziridine, 3-methyl-oxaziridine. 

\end{enumerate}

\section{Spectroscopic Databases for Interstellar Molecules}
In this paper, we commence our investigation with propylene oxide and other oxirane variants. In the course of trawling databases and the literature for suitable chiral candidates, though there is much astrochemical data available, we found no easy way to filter out the data for chiral molecules. We have thus made a start on a specialized database for interstellar chiral molecules (Section 4) intended to contain computational/experimental lab data for such molecules found in the ISM and potential targets. With only one chiral molecule having been detected in space the database is small but it is hoped that this will provide a ready reference in the quest for finding chirality in the interstellar medium and stellar accretion disks.  

The intention, as further described in \S4, is that the database will provide a useful start to an ongoing effort by the community. As an initial start, researchers can submit information to the authors to be included into the database but we hope that future collaboration with astrochemistry groups can help us build the database.

The notable databases already containing data on interstellar molecules (and see references within) are:

\begin{enumerate}

\item[i.] Splatalogue - \url{https://splatalogue.online/} \cite{SPLAT}\\

Splatalogue is the main database for observational astronomers, maintained by the National Radio Astronomy Observatory, but collating data from a number of laboratories worldwide. It contains over 5.8 million spectral lines in the radio to submillimeter range (400\,Hz to 3.22\,PHz).

\item[ii.] Computational Chemistry Comparison and Benchmark Database - \url{https://cccbdb.nist.gov/} \cite{nist}\\

CCCBDB is a database of experimental and computational quantum chemical data for a selection of gas-phase atoms and small molecules. It is aimed at chemists and provides the feature of being searchable by molecular formula.

\item[iii.]  Leiden Atomic and

Molecular Database - \url{https://home.strw.leidenuniv.nl/\~moldata/} \cite{Sho05} \\

The LAMDa database provides users of radiative transfer codes with the basic atomic and molecular data needed for excitation calculation. Currently, the database contains data for 47 molecular species.

\item[iv.] A Database of Molecules Detected in Space - \url{https://github.com/bmcguir2/astromol} \cite{McG22} \\

Astromol is a database of molecules detected in space and an object-oriented interface for interacting with the database. This additionally includes a {\sc Python 3} interface for making figures and tables for publications. The main focus of this database and included paper is on a ``Census of Interstellar, Circumstellar, Extragalactic, Protoplanetary Disk, and Exoplanetary Molecules."

\item[v.] High temperature molecular line lists for modelling exoplanet atmospheres - \url{https://www.exomol.com/} \cite{ExoMOL} \\

ExoMol is a list of molecules designed to help with modeling atmospheric conditions in exoplanets, brown dwarfs, and cool stars. In particular, the focus is on hot atmospheres.

\item[vi.] Cologne Database for Molecular Spectroscopy - \url{https://cdms.astro.uni-koeln.de/cdms/portal} \cite{CDMS}\\

CDMS includes molecules that may be contained within the interstellar or circumstellar medium or in planetary atmospheres. The catalog contains mostly rotational transition frequencies, uncertainties, intensities for radio and far infrared data. This information is often accessible through the $Splatalogue$ interface.

\item[vii.] Molecular Spectroscopy Jet Propulsion Laboratory - \url{https://spec.jpl.nasa.gov/} \\

The JPL catalog contains a collection of information helpful for spectral line fitting for radio and infrared data, including rotation constants and associated references for the work in which the information was determined. Some of this information is accessible through the $Splatalogue$ interface.

[\item[viii.] Molecular Gasphase Documentation - \url{https://www.uni-ulm.de/nawi/chemieinformationssysteme/mogadoc/introduction/}

MOGADOC is a database of numerical structural data such as internuclear distances, bond angles and dihedral angles. The database covers data from both laboratory and astronomical investigations.

\end{enumerate}

\section{Computational study of propylene oxide and oxirane variants}

Propylene oxide, being the first chiral molecule detected in space prompted \cite{Ell20} to consider its structural analogs as potential targets for radio astronomy search and \cite{Lankhaar_2022} to study the role magnetic fields may play in their creation. The structures \cite{Ell20} considered are given in Table~\ref{Table1}. As they point out they are not the lowest energy of the isomeric series but what is important in the interstellar medium is whether there is a path to their creation. Their formation may be a rare event but equally under those conditions once they form they are likely to persist. Ellinger {\it et al.} carried out density functional theory (DFT) calculations on these structures with triple-zeta quality basis sets. However, chemical accuracy requires higher-level methods so we have repeated their gas-phase calculations as described below.

\begin{table*}
\caption{Structures of propylene oxide and oxirane variants considered in this study.}
{\begin{tabular}{|p{1.5in}|p{1.5in}|p{1.5in}|}
 \hline
 Molecule & IUPAC name & Structure \\
 \hline
Propylene oxide & 2-methyloxirane &   \includegraphics{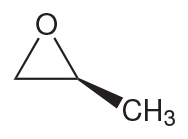}\\  
 Hydroxyoxirane & oxiran-2-ol &\includegraphics{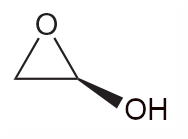} \\
 Cyanooxirane & oxirane-2-carbonitrile &\includegraphics{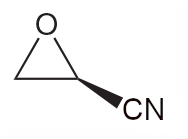} \\
 Aminooxirane & oxiran-2-amine &\includegraphics{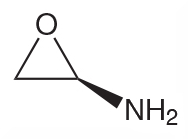} \\ 
 Ethynyloxirane & 2-ethynyloxirane
 &\includegraphics {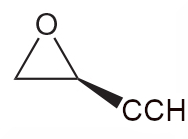} \\
 Formyloxirane & oxirane-2-carbaldehyde &\includegraphics{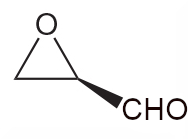} \\
 \hline
\end{tabular}}
\label{Table1}
\end{table*}

\subsection{Computational Details}
We followed the methodology used in our recent study of the isomers of HCCCCS \citep{Paper1} where it is explained in more detail. 

The structures for the molecules given in Table \ref{Table1} were initially optimized at CCSD(T) level \citep{CCSD1,CCSD2} with the cc-pVTZ \citep{cc-PVTZ} basis set using the Molpro quantum chemistry software package \citep{molpro}. This provides a good estimate of the structures but is not an adequate level for the prediction of rotational constants, so the optimization was refined using triple (TZ), quadruple (QZ), and quintuple (5Z) basis sets from the cc-pVnZ series \citep{cc-PVTZ}, together with corrections for core-correlation and relativistic effects.

$$E = E_{CBS-TQ5} + \Delta E_{core} +\Delta E_{rel}$$

The term E$_{CBS-TQ5}$ is the extrapolated energy calculated by using CCSD(T) calculations TZ, QZ, and 5Z basis sets and the extrapolation formula of \cite{ML96}. The other terms are those due to core correlation energy and relativistic effects respectively.

Due to constraints on the amount of memory that could be routinely used, the 5Z calculations had to be approximated slightly by only using QZ on the Hydrogen atoms, while retaining 5Z on the other atoms. The other terms in the energy expression require specialized basis sets, the Martin-Taylor basis \citep{MT94} for the core correlations term, and Douglas-Kroll basis set \citep{D-K,Jansen} for relativistic effects.

\subsection{Results and Discussion}

The molecular geometries were optimized using analytic gradients, assembled from the gradients of each term in the energy expression. Starting from the CCSD(T)/cc-pVTZ geometry, and using a hessian computed at that level, it took typically eight or nine iterations until the bond lengths were changing by about 0.0001\,{\AA}ngstrom, and the rotational constants changed by less than 1\,MHz. The resulting geometries are all given in Supplementary Information. All of the structures are local minima, {\it i.e.} have no negative eigenvalues of the force constant matrix.

Table \ref{Table2} contains experimental values of the rotational constants,  where available, and values calculated at equilibrium geometry. Various studies, for example, \cite{CS23}, indicate that rotational constants calculated at an accurate equilibrium geometry will be larger than the true ground state values by about 1\,percent, or less. This is the case here. However, an inaccuracy of 1\,percent is not acceptable for spectroscopic purposes, where \cite{CS23} suggest an error of 0.1\,percent, or less, should be the aim.

In order to improve upon the approximately 99\,percent accuracy obtained just from using equilibrium geometry, it is necessary to allow for vibrational coupling. Very high accuracy can be achieved by calculating the energy on a multidimensional grid of the geometric coordinates, fitting the energies to an analytic representation of the potential energy surface, and calculating the vibration-rotation coupling with a variation in vibrational SCF/CI program ({\it e.g.} \cite{Lee2}). However, this is extremely demanding computationally and is only practical for small molecules with three or four atoms. An alternative approach, more practical for larger molecules, is to use perturbation theory in conjunction with a more approximate cubic or quartic force field \citep{CS23, CP17}. A similar approach is used here, specifically using the second-order vibrational perturbation theory program of {\cite{VPT2}, which has been incorporated into the Gaussian software suite \citep{G16}, together with the double-hybrid B2PLYP method of \cite{B2PLYP} to produce the necessary anharmonic force constants. For the harmonic force constants CCSD(T)/cc-pVTZ values were used. For the anharmonic terms an aug-cc-pVTZ basis \citep{Kendall} was used. The results are in Table \ref{Table2}. In all cases, the agreement with experimental values is improved, and in the majority of cases, the deviation from the experiment is now less than 0.1\,percent. The worst match to the experiment comes from oxirane-2-carbaldehyde (formyloxirane). The experimental values for this are for the trans-isomer, as are our calculations, but there is a cis-isomer \citep{RBG10} which means that vibrational averaging may need to be more sophisticated than with just perturbation theory.

Calculated and experimental dipole moments are given in Table \ref{Table3}. The Cartesian components of the dipole are included in the Supplementary Information attached to this paper. We note that most of the molecules we investigated have values that were already published, so they provide a basis for understanding the accuracy of the new values produced for hydroxyoxirane and aminooxirane.

\begin{table*}
\raggedright
\caption{Rotational constants A,B,C in GHz from experiment, for the calculated equilibrium geometry and for the calculated ground vibrational state. The commonest isotope masses have been used i.e 1-H,12-C,14-N and 16-O. The values in brackets are the percentage deviation from the experimental values, where available.}
{\begin{tabular}{p{1.2in} p{1.2in} p{1.85in} p{1.2in} p{1.2in}}

 \hline
 \multicolumn{1}{c}{Molecule} & \multicolumn{1}{c}{Experimental} & \multicolumn{1}{c}{Reference}& \multicolumn{1}{c}{equilibrium} & \multicolumn{1}{c}{ground state}\\
 \hline
Propylene oxide & 18.023845,6.682369,& \cite{Mesko}& 18.1896,6.7412,6.0065& 18.0321,6.6781,5.9486\\ 
&5.951176 &  &(0.920,0.885,0.926)& (0.047,-0.059,-0.046)\\
 Hydroxyoxirane & none &&19.8597,7.4035,6.4323 &19.6629,7.3516,6.3856 \\
 Cyanooxirane &18.456295,3.524705,&\cite{MB96} & 18.6063,3.5357,3.3799& 18.4906,3.5234,3.3661\\
 & 3.367811 & &(0.812,0.311,0.358) & (0.186,-0.037,-0.051) \\
 Aminooxirane & none&&19.3292,7.1287,6.2082& 19.1624,7.0699,6.1619\\ 
 Ethynyloxirane & 18.41197,3.41562, &\cite{CMAS77}& 18.5577,3.4262,3.2820&18.4422,3.4149,3.2695\\
 & 3.27074& &(0.791,0.310,0.344) & ( 0.164,-0.021,-0.038) \\
 Formyloxirane &18.24135,3.27289, &\cite{CB03}&18.4041,3.2889,3.1572 &18.1759,3.2785,3.1399\\
 & 3.13762& &(0.892,0.489,0.624) & ( -0.362,0.150,0.060) \\
 \hline
 \end{tabular}}
\label{Table2}
\end{table*}

\begin{table*}
\raggedright
\caption{Experimental and calculated Dipole moments, in Debye}
{\begin{tabular}{lclc}
 \hline
 \multicolumn{1}{c}{Molecule} & \multicolumn{1}{c}{Experimental} & \multicolumn{1}{c}{Reference} & \multicolumn{1}{c}{Calculated} \\
 \hline
Propylene oxide & 2.001&\cite{CS77} & 2.001\\ 
 Hydroxyoxirane & none && 1.492 \\
 Cyanooxirane &3.740&\cite{MB96} & 3.745\\
 Aminooxirane & none&& 3.287\\ 
 Ethynyloxirane & 1.795&\cite{CMAS77} & 1.850\\
 Formyloxirane & 2.469&\cite{CB03} & 2.534\\
 \hline
 \end{tabular}}
\label{Table3}
\end{table*}

Calculated centrifugal distortion constants for propylene oxide are given in Table \ref{Table4}, and are in surprisingly good agreement with experimental data.
The equivalent data for the other molecules are included in the Supplementary Information.

\begin{table*}
\caption{Centrifugal distortion constants for propylene oxide, in kHz}
{\begin{tabular}{lcc}
 \hline
 \multicolumn{1}{c}{Molecule} & \multicolumn{1}{c}{Experimental} & \multicolumn{1}{c}{Calculated} \\
 &\citep{Mesko}&\\
 \hline
$\Delta_J$ & 2.9146 & 2.977\\ 
 $\Delta_{JK}$ & 3.4672 & 3.263 \\
 $\Delta_K$ & 19.7271 & 19.708\\
$\delta_J$ & 0.1929& 0.186\\ 
 $\delta_K$ & 2.6004 & 2.391\\
 \hline
 \end{tabular}}
\label{Table4}
\end{table*}

\section{The CHARISMA database for interstellar chiral molecules}

As mentioned in Section 2, though there exist many databases in astronomy and chemistry containing experimental and computational data to aid in the identification of molecules detected in the interstellar medium, it is very difficult to filter out the results for chiral molecules. To this end, we have begun to put together a database, whimsically called CHARISMA (Chiral Assay of Radio ISM Astrophysics), specifically dedicated to chiral molecules found and potentially found in the ISM. This is presented as a fledgling proof-of-concept database and it is hoped that the community will find it useful enough to maintain. The database is based on the MolDB6 software package of Norbert Haider \citep{MolDB6} and can be found at the URL~\url{http://chiral-ISMDB.net/}. The interface as shown in Fig.~\ref{fig:ISMDB1} can allow searching by name or structure and example search results found searching for oxirane by name or structure are shown in Fig.~\ref{fig:ISMDB2}.

\begin{figure*}
\begin{tabular}{l}
(a) Search by name\\
\includegraphics[width=0.94\textwidth]{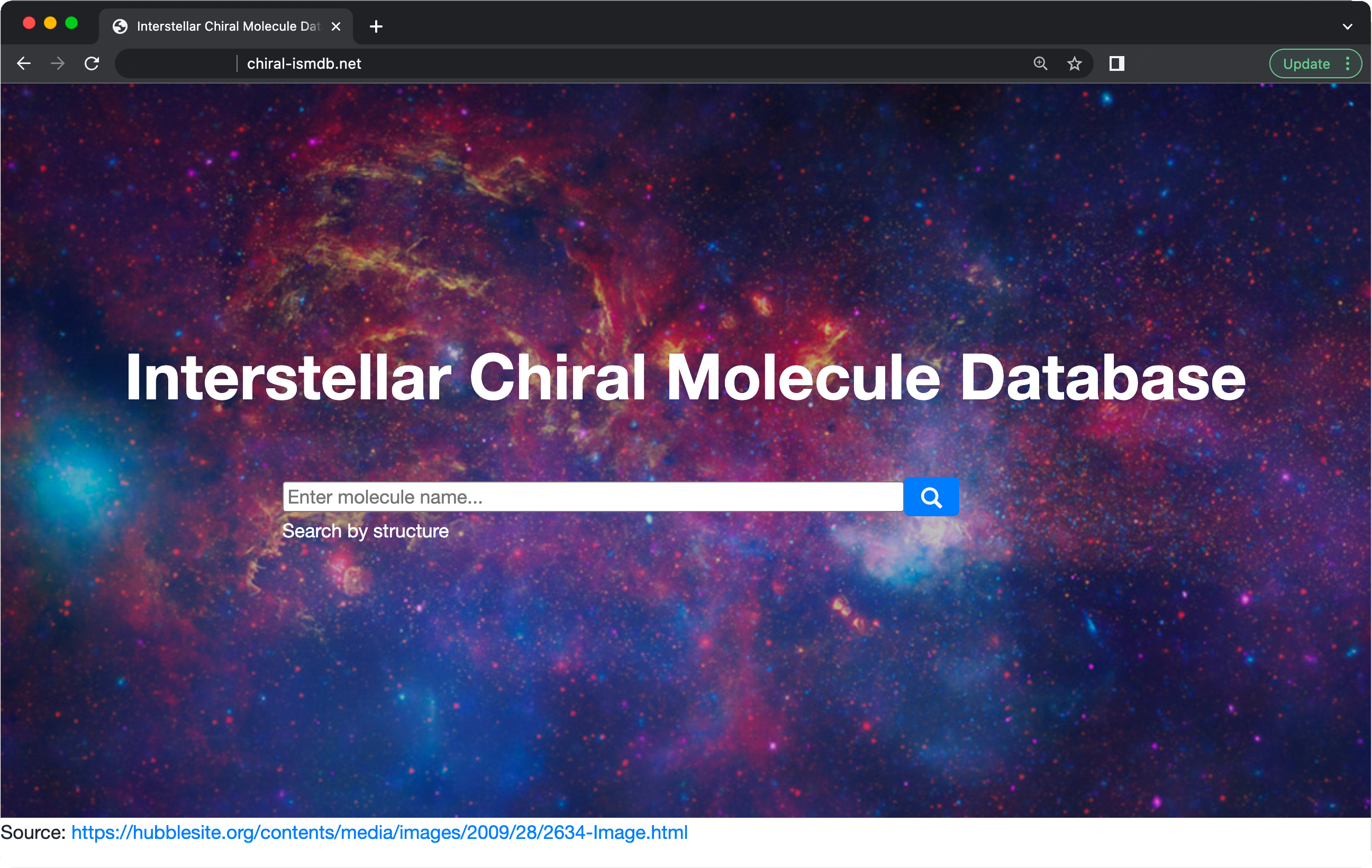} \\
(b) Search by structure \\
\includegraphics[width=0.94\textwidth]{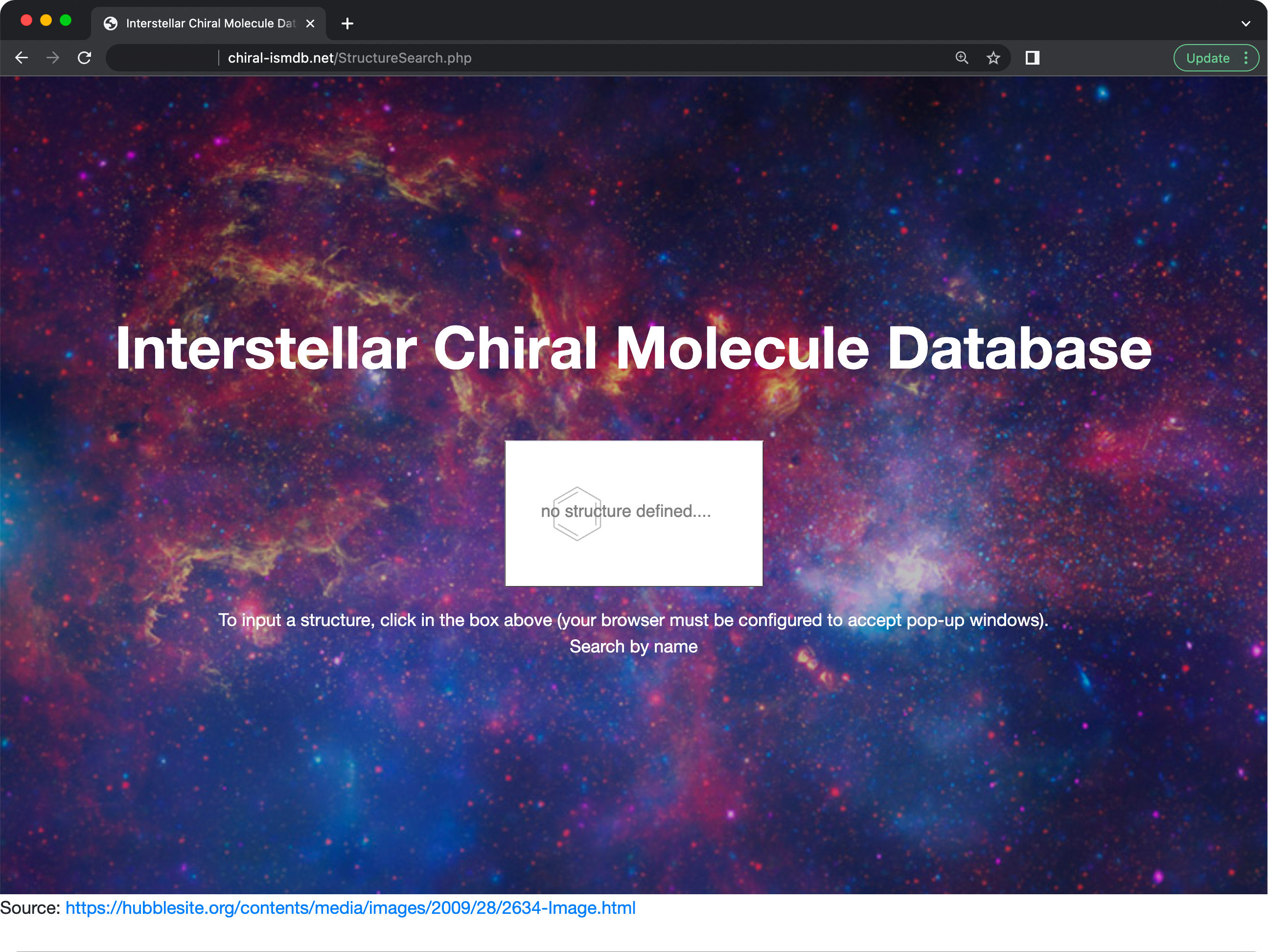}
\end{tabular}
\caption{Screenshots of the CHARISMA search page (a) by name and (b) by structure. \label{fig:ISMDB1}}
\end{figure*}

\begin{figure*}
\includegraphics[width=0.97\textwidth]{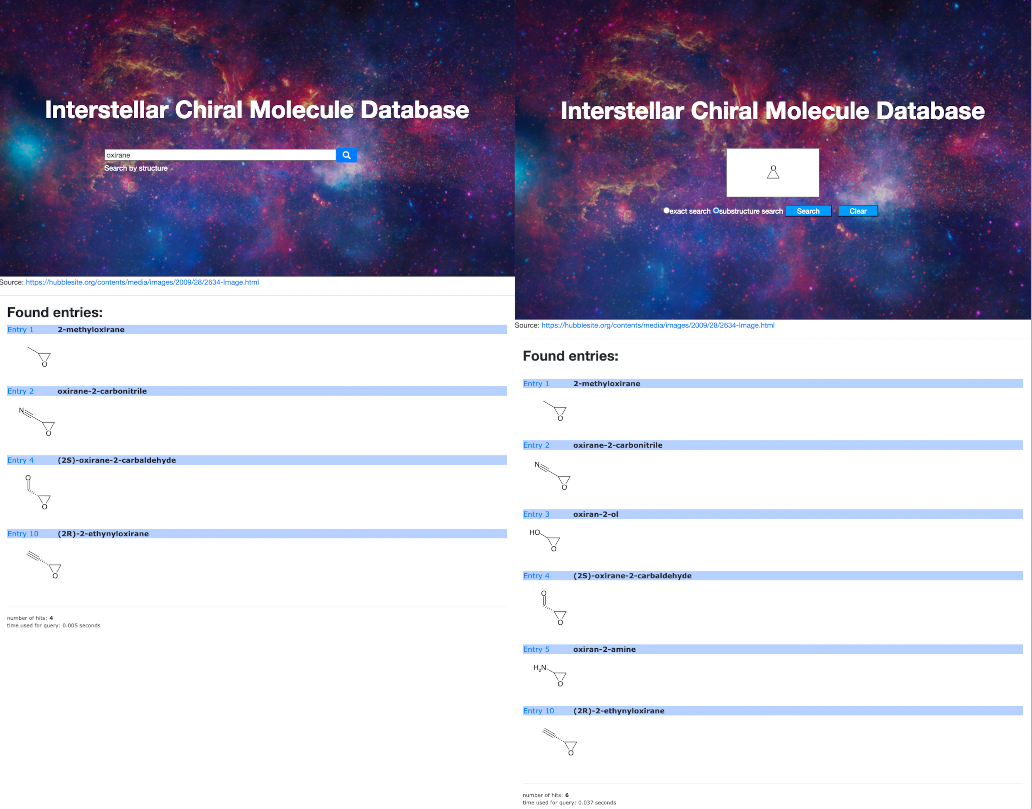}
\caption{Screenshots of the CHARISMA search results. \label{fig:ISMDB2}}
\end{figure*}

Note that they do not return the same number of hits as not all the compounds contained oxirane in their name {\it e.g.} oxiran-2-ol. This was a problem we found searching for spectroscopic data on molecules of interest where there was inconsistency in nomenclature. An example of a results page, that for propylene oxide itself, is given in Fig.~\ref{fig:ISMDB3}. It provides basic identification information, a description relating it to astronomical discovery and experimental/calculated rotational constants where available. With feedback from the community more fields of interest can be added.

\begin{figure*}
\includegraphics[width=0.9\textwidth]{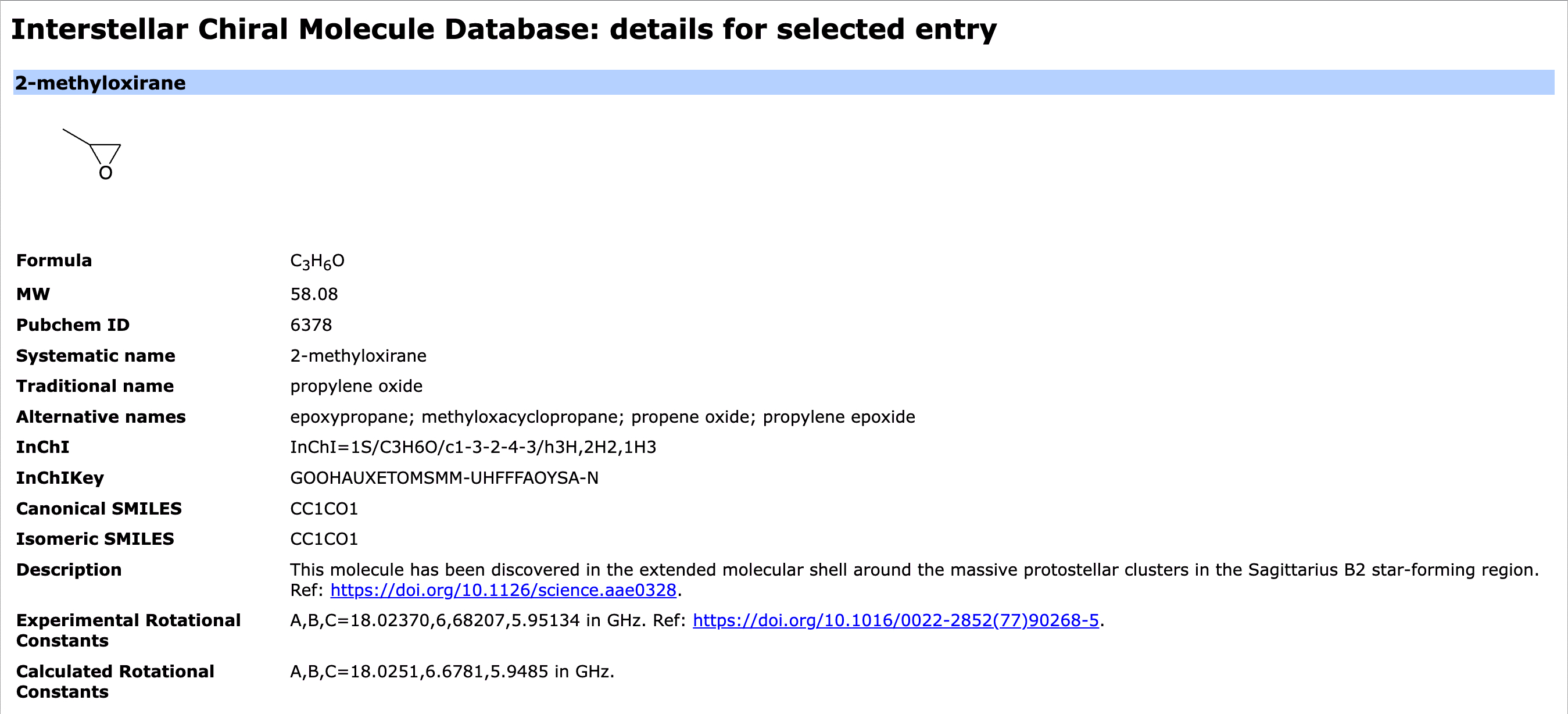}
\caption{Screenshot of the propylene oxide entry from the CHARISMA database. \label{fig:ISMDB3}}
\end{figure*}

\section{Future outlook}

Understanding the chemical origins of life is a primary science goal of many of the upcoming and new telescopes around the world. Among the many challenges and unanswered questions is understanding the homochirality puzzle; that life on Earth has a preference for left-hand chirality. In particular, the question remains if homochirality, which allows biopolymers to adopt helical structures, is imposed from astronphysical origins. 

The strategies for determining the answer to this question vary from the theoretical approaches on how homochirality could come about, as presented by \cite{Glo21}, models for understanding dichroic activity in the interstellar medium \citep{Lankhaar_2022}, to observational astronomy \citep{McG22}. However, despite this work, still little is understood of the nature of chiral molecules in the Universe. In order to approach the problem from observational astronomy, the process is expedited through knowledge of the most likely transitions to be detectable in the interstellar medium, the frequency of this emission, the type of transition (ro-vibrational, vibrational, or rotational), when rotational, the dipole moment, and values for the rotational partition function as per environment temperatures. Some of this information is provided in a series of online catalogs, but a database with a focus on chiral molecules and their expected precursors would significantly focus our search. 

Once the list of molecules, information that leads to understanding which transitions are most likely detectable, and the frequencies at which the search should be focused are collected in a single location, the observational search can begin in a more focused manner. Although rapid computational analysis such as presented by \cite{Zapata_2021} or DFT analysis by \cite{Alessandrini2023} provides a quick analysis of frequencies that provide interesting results, analysis such as provided in this paper with the 99.9\,percent accuracy (preferably with multiple transitions in a given band of the spectrum) is critical in the true identification of molecules, especially in places where a significant number of molecular transitions can exist (line forests). Even with this high accuracy, it is through the detection of multiple transitions with the expected frequency separation, in which a true detection would be made.

In work by \cite{McG22}, he showed that out of the 241 molecules so far detected in astronomical sources, the majority of the largest molecular species were detected at cm wavelengths. Also, for future discoveries, the probability detection of larger molecular species (molecular structures containing more atoms) is favored toward centimeter or millimeter wavelengths. Work by \cite{Tre20} suggested that molecular structures with larger atoms may have favored transitions toward meter wavelengths, including molecular isotopologues. Therefore a focus on information about transitions observable between 700\,MHz -- 25\,GHz, would provide a significant benefit to the astronomical community.

%%\section{Summary and conclusions}
%%\label{}
%%\lipsum[1-4]
\section{Conclusion}
In this paper, we present accurate analysis for oxirane chiral variants with a 99.9\,percent accuracy in laboratory experiments.  As this information for important chiral molecules, their transitions, and relevant information becomes available, we have created a database (CHARISMA) specifically designed to house this information.  The plan is to make this work a stepping point in building the necessary information needed to utilize the most sensitive radio telescopes being built to discover our cosmic origins if they exist.

\section*{Acknowledgments}
     The calculations were carried out on the NCI National Facility in Canberra, Australia, which is supported by the Australian Commonwealth Government.
     RK would like to thank Claire Trenham, David Monro, and Andrew Robinson for advice on data collection software, and Professor Norbert Haider for advice on his MolDB6 software package.\\

\appendix

\section{Supplementary Material}

\begin{table}
\caption{Optimized geometry of Oxi-OH in Angstrom and calculated dipole, in Debyes. }
{\begin{tabular}{lccc}
 \hline
 \multicolumn{1}{c}{} &\multicolumn{1}{c}{X} & \multicolumn{1}{c}{Y} & \multicolumn{1}{c}{Z} \\
 
 \hline
  C &       --0.98876108   &   0.60999214  &   -0.03171518\\
  C        &        0.21044540   &  -0.04437814     & 0.45509070\\
  O        &       --0.75336557   &  -0.79717721    & -0.25103706\\
  H       &        --1.80344274   &   0.83798169   &   0.64358301\\
  H       &        -0.89849029 &  7   1.22120212   &  --0.92049251\\
  H       &         0.31405706   &  --0.32524228   &   1.49832614\\
  O       &         1.39108931   &   0.23118277   &  --0.19717210\\
  H       &         1.96790390   &  --0.53435410  &   --0.12234701\\
Dipole        &     0.4707        &  0.3324     &     1.3761\\
 \hline
 \end{tabular}}
\end{table}  

\begin{table}[]
\caption{Optimized geometry of Oxi-NH2, in Angstrom, and calculated dipole in Debyes. }
{\begin{tabular}{lccc}
 \hline
 \multicolumn{1}{c}{} &\multicolumn{1}{c}{X} & \multicolumn{1}{c}{Y} & \multicolumn{1}{c}{Z} \\
 
 \hline
  C       &        --0.99312421  &    0.63175994     &--0.03052354\\
  C      &          0.19146569  &   --0.06462589    &  0.46038256\\
  O       &        --0.81294815   &  --0.77069000    & --0.25159367\\
  H       &        --1.79226903   &   0.89029288    &  0.65325734\\
  H       &        --0.88994716   &   1.24150787    & --0.91969613\\
  H       &         0.22919750   &  --0.34621549    &  1.50998755\\
  N       &         1.42041767   &   0.11706126    & --0.23771989\\
  H       &         1.85988743   &   0.98706165   &   0.03345766\\
  H       &         2.06678024   &  --0.63597718   &  --0.04228488\\
Dipole      &       1.2543   &       1.6724     &     1.8445\\
 \hline
 \end{tabular}}
\end{table} 

\begin{table}
\caption{Optimized geometry of Oxi-CH3, in Angstrom, and calculated dipole in Debyes. }
{\begin{tabular}{lccc}
 \hline
 \multicolumn{1}{c}{} &\multicolumn{1}{c}{X} & \multicolumn{1}{c}{Y} & \multicolumn{1}{c}{Z} \\
 
 \hline
   C        &       --1.02942728 &     0.61851566  &   --0.06001151\\
  C       &         0.15448626   &  --0.03517206   &   0.49293709\\
  O       &        --0.81850822   &  --0.78613487   &  -0.23846862\\
  H       &        --1.85928376   &   0.87556365   &   0.58710804\\
  H       &        --0.92665400  &    1.21012047   &  --0.96273565\\
  H       &         0.14984783  &   --0.25465642   &   1.55662781\\
  C       &         1.50041895  &    0.10014683   &  --0.14846823\\
  H       &         2.06453790  &    0.90887760   &   0.31813194\\
  H       &         2.07100298  &   --0.82204288    & --0.03852508\\
  H        &        1.39001736  &    0.31412498    & -1.21069982\\
Dipole     &        0.8453     &     1.6844    &      0.6894\\
 \hline
 \end{tabular}}
\end{table} 

\begin{table}
\caption{Optimised geometry of Oxi-CN in Angstrom, and calculated dipole in Debyes. }
{\begin{tabular}{lccc}
 \hline
 \multicolumn{1}{c}{} &\multicolumn{1}{c}{X} & \multicolumn{1}{c}{Y} & \multicolumn{1}{c}{Z} \\
 
 \hline
     C      &          0.02373255  &    0.00621818    & -0.03619446\\
  C        &       --0.03564931   &   0.04285386    &  1.42891457\\
  O       &         1.21308344  &    0.02391650   &   0.74963117\\
  H       &        --0.18364573  &   -0.93072925   &  -0.53684727\\
  H       &        --0.17350203  &    0.91664362   &  -0.58734822\\
  H       &        --0.27880665  &   --0.85652277   &   1.98021670\\
  C       &        --0.37255724  &    1.28117791   &   2.10502155\\
  N       &        --0.66313701  &    2.25856500   &   2.64972190\\
Dipole    &        --0.9282     &    --2.9737   &      --2.1313\\
 \hline
 \end{tabular}}
\end{table} 

\begin{table}
\caption{Oxi-CCH optimized geometry in Angstrom, and calculated dipole, in Debyes. }
{\begin{tabular}{lccc}
 \hline
 \multicolumn{1}{c}{} &\multicolumn{1}{c}{X} & \multicolumn{1}{c}{Y} & \multicolumn{1}{c}{Z} \\
 
 \hline
  C        &       -1.34360189 &     0.67731116  &   -0.05347865\\
  C        &       -0.29711442  &   -0.16017985   &   0.54273030\\
  O        &       -1.20010069  &   -0.68906701   &  -0.43335326\\
  H        &       -2.23065641 &     0.90263220   &   0.52854566\\
  H         &      -1.06063886  &    1.39532373   &  -0.81500970\\
  H        &       -0.45468044  &   -0.54592298   &   1.54457311\\
  C        &        1.08186687  &   -0.01921482  &    0.14111340\\
  C        &        2.23920952  &    0.10100887  &   -0.17083024\\
  H        &        3.25935131  &    0.20217166   &  -0.45385563\\
Dipole     &        0.3005     &     1.5296     &     0.9959\\
 \hline
 \end{tabular}}
\end{table} 

\begin{table}
\caption{Oxi-HCO optimised geometry in Angstrom, and calculated dipole in Debyes. }
{\begin{tabular}{lccc}
 \hline
 \multicolumn{1}{c}{} &\multicolumn{1}{c}{X} & \multicolumn{1}{c}{Y} & \multicolumn{1}{c}{Z} \\
 
 \hline
  C       &        --1.40544299   &   0.65271721    & --0.08525241\\
  C       &        --0.22808739   &  --0.02889079    &  0.47018990\\
  O       &        --1.31312217   &  --0.76350038    & -0.10192572\\
  H       &        --2.13423288   &   1.06393872    &  0.60391230\\
  H       &        --1.33473720   &   1.12851209    & -1.05856862\\
  H       &        --0.10621023   &  --0.11464224    &  1.54469015\\
  C       &         1.01452697   &  --0.04545561    & -0.34712604\\
  O       &         2.11761213   &   0.09330637     & 0.11940617\\
  H       &         0.84636473   &  --0.19056238     &--1.43188971\\
Dipole    &        --2.0439      &    1.4744      &   --0.2708\\
 \hline
 \end{tabular}}
\end{table}

\begin{table*}
\caption{Centrifugal distortion constants, in kHz. Experimental values from Muller(1996) and Collins(1972).  }
{\begin{tabular}{lcccc}
 \hline
 \multicolumn{1}{c}{}&\multicolumn{1}{c}{Oxi-HCO, expt} &\multicolumn{1}{c}{Oxi-HCO,calc} & \multicolumn{1}{c}{Oxi-CN,expt} & \multicolumn{1}{c}{Oxi-CN,calc} \\
 
 \hline	
$\Delta_J$&	0.5735&0.5761&	1.1472&1.4414\\
$\Delta_{JK}$&	12.3899&	8.4736&	-1.7094&	-4.2772\\
$\Delta_K$&17.2941&	28.7092&	75.858&	107.037\\
$\delta_J$&0.02388&	0.03063&	-0.0495&	-0.0854\\
$\delta_K$&3.1171&	9.8561&	1.392&	1.4525\\
 \hline
 \end{tabular}}
\end{table*} 

\begin{table*}
\caption{Calculated centrifugal distortion constants, in kHz.  There are no experimental data for these systems.  }
{\begin{tabular}{lccc}
 \hline
 \multicolumn{1}{c}{}&\multicolumn{1}{c}{Oxi-OH,calc} &\multicolumn{1}{c}{Oxi-NH2,calc} & \multicolumn{1}{c}{Oxi-CCH,calc}  \\
 
 \hline	

$\Delta_J$&	3.2323&3.0066&	1.2402\\
$\Delta_{JK}$&	5.3286&	7.3518&	-4.2979\\
$\Delta_K$&26.4109&	19.9123&	118.6536\\
$\delta_J$&0.3308&	0.2726&	-0.0897\\
$\delta_K$&	5.3811&	2.3274&	1.0932\\
 \hline
 \end{tabular}}
\end{table*} 
\bibliographystyle{elsarticle-harv} 
\bibliography{Chiral}

\begin{thebibliography}{61}
\expandafter\ifx\csname natexlab\endcsname\relax\def\natexlab#1{#1}\fi
\providecommand{\url}[1]{\texttt{#1}}
\providecommand{\href}[2]{#2}
\providecommand{\path}[1]{#1}
\providecommand{\DOIprefix}{doi:}
\providecommand{\ArXivprefix}{arXiv:}
\providecommand{\URLprefix}{URL: }
\providecommand{\Pubmedprefix}{pmid:}
\providecommand{\doi}[1]{\href{http://dx.doi.org/#1}{\path{#1}}}
\providecommand{\Pubmed}[1]{\href{pmid:#1}{\path{#1}}}
\providecommand{\bibinfo}[2]{#2}
\ifx\xfnm\relax \def\xfnm[#1]{\unskip,\space#1}\fi
%Type = Article
\bibitem[{{Alessandrini} et~al.(2023){Alessandrini}, {Melosso}, {Rivilla},
  {Bizzocchi} and {Puzzarini}}]{Alessandrini2023}
\bibinfo{author}{{Alessandrini}, S.}, \bibinfo{author}{{Melosso}, M.},
  \bibinfo{author}{{Rivilla}, V.M.}, \bibinfo{author}{{Bizzocchi}, L.},
  \bibinfo{author}{{Puzzarini}, C.}, \bibinfo{year}{2023}.
\newblock \bibinfo{title}{{Computational Protocol for the Identification of
  Candidates for Radioastronomical Detection and Its Application to the C3H3NO
  Family of Isomers}}.
\newblock \bibinfo{journal}{Molecules} \bibinfo{volume}{28},
  \bibinfo{pages}{3226}.
\newblock \DOIprefix\doi{10.3390/molecules28073226}.
%Type = Article
\bibitem[{Amos and Kobayashi(2023)}]{Paper1}
\bibinfo{author}{Amos, R.D.}, \bibinfo{author}{Kobayashi, R.},
  \bibinfo{year}{2023}.
\newblock \bibinfo{title}{Computational studies of hccccs isomers}.
\newblock \bibinfo{journal}{Molecular Physics} \bibinfo{volume}{0},
  \bibinfo{pages}{e2203269}.
\newblock \URLprefix \url{https://doi.org/10.1080/00268976.2023.2203269},
  \DOIprefix\doi{10.1080/00268976.2023.2203269}.
%Type = Article
\bibitem[{{Barone} et~al.(2014){Barone}, {Biczysko}, {Bloino} and
  {Puzzarini}}]{BBBP14}
\bibinfo{author}{{Barone}, V.}, \bibinfo{author}{{Biczysko}, M.},
  \bibinfo{author}{{Bloino}, J.}, \bibinfo{author}{{Puzzarini}, C.},
  \bibinfo{year}{2014}.
\newblock \bibinfo{title}{{Accurate molecular structures and infrared spectra
  of trans-2,3-dideuterooxirane, methyloxirane, and
  trans-2,3-dimethyloxirane}}.
\newblock \bibinfo{journal}{JCP} \bibinfo{volume}{141},
  \bibinfo{pages}{034107}.
\newblock \DOIprefix\doi{10.1063/1.4887357}.
%Type = Article
\bibitem[{{Barone} et~al.(2005){Barone}, {Carbonniere} and {Pouchan}}]{VPT2}
\bibinfo{author}{{Barone}, V.}, \bibinfo{author}{{Carbonniere}, P.},
  \bibinfo{author}{{Pouchan}, C.}, \bibinfo{year}{2005}.
\newblock \bibinfo{title}{{Accurate vibrational spectra and magnetic properties
  of organic free radicals: The case of H$_{2}$CN}}.
\newblock \bibinfo{journal}{Journal of Chemical Physics} \bibinfo{volume}{122},
  \bibinfo{pages}{224308--224308}.
\newblock \DOIprefix\doi{10.1063/1.1924592}.
%Type = Article
\bibitem[{{Belloche} et~al.(2008){Belloche}, {Menten}, {Comito}, {M{\"u}ller},
  {Schilke}, {Ott}, {Thorwirth} and {Hieret}}]{Bel08}
\bibinfo{author}{{Belloche}, A.}, \bibinfo{author}{{Menten}, K.M.},
  \bibinfo{author}{{Comito}, C.}, \bibinfo{author}{{M{\"u}ller}, H.S.P.},
  \bibinfo{author}{{Schilke}, P.}, \bibinfo{author}{{Ott}, J.},
  \bibinfo{author}{{Thorwirth}, S.}, \bibinfo{author}{{Hieret}, C.},
  \bibinfo{year}{2008}.
\newblock \bibinfo{title}{{Detection of amino acetonitrile in Sgr B2(N)}}.
\newblock \bibinfo{journal}{AAP} \bibinfo{volume}{482},
  \bibinfo{pages}{179--196}.
\newblock \DOIprefix\doi{10.1051/0004-6361:20079203},
  \href{http://arxiv.org/abs/0801.3219}{{\tt arXiv:0801.3219}}.
%Type = Misc
\bibitem[{{Blackmond}(2019)}]{cshlpOriginBiological}
\bibinfo{author}{{Blackmond}, D.}, \bibinfo{year}{2019}.
\newblock \bibinfo{title}{{T}he {O}rigin of {B}iological {H}omochirality ---
  cshperspectives.cshlp.org}.
\newblock
  \bibinfo{howpublished}{\url{https://cshperspectives.cshlp.org/content/11/3/a032540.abstract}}.
\newblock \bibinfo{note}{[Accessed 20-Jul-2023]}.
%Type = Article
\bibitem[{{Chen} and {Ma}(2020)}]{CM20}
\bibinfo{author}{{Chen}, Y.}, \bibinfo{author}{{Ma}, W.}, \bibinfo{year}{2020}.
\newblock \bibinfo{title}{{The origin of biological homochirality along with
  the origin of life}}.
\newblock \bibinfo{journal}{PLoS Computational Biology} \bibinfo{volume}{16},
  \bibinfo{pages}{e1007592}.
\newblock \DOIprefix\doi{10.1371/journal.pcbi.1007592}.
%Type = Article
\bibitem[{{Collins} and {Boggs}(1972)}]{CB03}
\bibinfo{author}{{Collins}, M.J.}, \bibinfo{author}{{Boggs}, J.E.},
  \bibinfo{year}{1972}.
\newblock \bibinfo{title}{{Microwave Spectrum, Structure, and Dipole Moment of
  Epoxybutyne}}.
\newblock \bibinfo{journal}{JCP} \bibinfo{volume}{57},
  \bibinfo{pages}{3811--3815}.
\newblock \DOIprefix\doi{10.1063/1.1678849}.
%Type = Article
\bibitem[{{Cooper} et~al.(2001){Cooper}, {Kimmich}, {Belisle}, {Sarinana},
  {Brabham} and {Garrel}}]{Coop21}
\bibinfo{author}{{Cooper}, G.}, \bibinfo{author}{{Kimmich}, N.},
  \bibinfo{author}{{Belisle}, W.}, \bibinfo{author}{{Sarinana}, J.},
  \bibinfo{author}{{Brabham}, K.}, \bibinfo{author}{{Garrel}, L.},
  \bibinfo{year}{2001}.
\newblock \bibinfo{title}{{Carbonaceous meteorites as a source of sugar-related
  organic compounds for the early Earth}}.
\newblock \bibinfo{journal}{Nat} \bibinfo{volume}{414},
  \bibinfo{pages}{879--883}.
\newblock \DOIprefix\doi{10.1038/414879A}.
%Type = Article
\bibitem[{{Coutens} et~al.(2022){Coutens}, {Loison}, {Boulanger}, {Caux},
  {M{\"u}ller}, {Wakelam}, {Manigand} and {J{\o}rgensen}}]{Coutens2022}
\bibinfo{author}{{Coutens}, A.}, \bibinfo{author}{{Loison}, J.C.},
  \bibinfo{author}{{Boulanger}, A.}, \bibinfo{author}{{Caux}, E.},
  \bibinfo{author}{{M{\"u}ller}, H.S.P.}, \bibinfo{author}{{Wakelam}, V.},
  \bibinfo{author}{{Manigand}, S.}, \bibinfo{author}{{J{\o}rgensen}, J.K.},
  \bibinfo{year}{2022}.
\newblock \bibinfo{title}{{The ALMA-PILS survey: First tentative detection of
  3-hydroxypropenal (HOCHCHCHO) in the interstellar medium and chemical
  modeling of the C$_{3}$H$_{4}$O$_{2}$ isomers}}.
\newblock \bibinfo{journal}{AAP} \bibinfo{volume}{660}, \bibinfo{pages}{L6}.
\newblock \DOIprefix\doi{10.1051/0004-6361/202243038},
  \href{http://arxiv.org/abs/2203.14119}{{\tt arXiv:2203.14119}}.
%Type = Article
\bibitem[{{Creswell} et~al.(1977){Creswell}, {Manor}, {Assink} and
  {Schwendeman}}]{CMAS77}
\bibinfo{author}{{Creswell}, R.A.}, \bibinfo{author}{{Manor}, P.J.},
  \bibinfo{author}{{Assink}, R.A.}, \bibinfo{author}{{Schwendeman}, R.H.},
  \bibinfo{year}{1977}.
\newblock \bibinfo{title}{{Microwave spectrum, torsional excitation energy,
  partial structure, and dipole moment of oxiranecarboxaldehyde}}.
\newblock \bibinfo{journal}{Journal of Molecular Spectroscopy}
  \bibinfo{volume}{64}, \bibinfo{pages}{365--375}.
\newblock \DOIprefix\doi{10.1016/0022-2852(77)90222-3}.
%Type = Article
\bibitem[{{Creswell} and {Schwendeman}(1977)}]{CS77}
\bibinfo{author}{{Creswell}, R.A.}, \bibinfo{author}{{Schwendeman}, R.H.},
  \bibinfo{year}{1977}.
\newblock \bibinfo{title}{{Centrifugal distortion constants and structural
  parameters of methyl oxirane}}.
\newblock \bibinfo{journal}{Journal of Molecular Spectroscopy}
  \bibinfo{volume}{64}, \bibinfo{pages}{295--301}.
\newblock \DOIprefix\doi{10.1016/0022-2852(77)90268-5}.
%Type = Inproceedings
\bibitem[{{De Pree} et~al.(2021){De Pree}, {Wilner}, {Kristensen},
  {Galvan-Madrid}, {Goss}, {Klessen}, {Mac Low}, {Peters}, {Robinson}, {Sloman}
  and {Rao}}]{De21}
\bibinfo{author}{{De Pree}, C.G.}, \bibinfo{author}{{Wilner}, D.},
  \bibinfo{author}{{Kristensen}, L.}, \bibinfo{author}{{Galvan-Madrid}, R.},
  \bibinfo{author}{{Goss}, M.}, \bibinfo{author}{{Klessen}, R.},
  \bibinfo{author}{{Mac Low}, M.}, \bibinfo{author}{{Peters}, T.},
  \bibinfo{author}{{Robinson}, A.}, \bibinfo{author}{{Sloman}, S.},
  \bibinfo{author}{{Rao}, M.}, \bibinfo{year}{2021}.
\newblock \bibinfo{title}{{Time-Variable Radio Recombination Line Emission in
  W49A}}, in: \bibinfo{booktitle}{American Astronomical Society Meeting
  Abstracts}, p. \bibinfo{pages}{114.07}.
%Type = Article
\bibitem[{{Dickens} et~al.(1996){Dickens}, {Irvine}, {Ohishi}, {Arrhenius},
  {Pitsch}, {Bauder}, {M{\"u}ller} and {Eschenmoser}}]{Dickens1996}
\bibinfo{author}{{Dickens}, J.E.}, \bibinfo{author}{{Irvine}, W.M.},
  \bibinfo{author}{{Ohishi}, M.}, \bibinfo{author}{{Arrhenius}, G.},
  \bibinfo{author}{{Pitsch}, S.}, \bibinfo{author}{{Bauder}, A.},
  \bibinfo{author}{{M{\"u}ller}, F.}, \bibinfo{author}{{Eschenmoser}, A.},
  \bibinfo{year}{1996}.
\newblock \bibinfo{title}{{A search for interstellar oxiranecarbonitrile
  (C$_{3}$H$_{3}$NO)}}.
\newblock \bibinfo{journal}{Origins of Life and Evolution of the Biosphere}
  \bibinfo{volume}{26}, \bibinfo{pages}{97--110}.
\newblock \DOIprefix\doi{10.1007/BF01809850}.
%Type = Article
\bibitem[{{Douglas} and {Kroll}(1974)}]{D-K}
\bibinfo{author}{{Douglas}, M.}, \bibinfo{author}{{Kroll}, N.M.},
  \bibinfo{year}{1974}.
\newblock \bibinfo{title}{{Quantum electrodynamical corrections to the fine
  structure of helium}}.
\newblock \bibinfo{journal}{Annals of Physics} \bibinfo{volume}{82},
  \bibinfo{pages}{89--155}.
\newblock \DOIprefix\doi{10.1016/0003-4916(74)90333-9}.
%Type = Article
\bibitem[{{Dunning}(1989)}]{cc-PVTZ}
\bibinfo{author}{{Dunning}, Thom~H., J.}, \bibinfo{year}{1989}.
\newblock \bibinfo{title}{{Gaussian basis sets for use in correlated molecular
  calculations. I. The atoms boron through neon and hydrogen}}.
\newblock \bibinfo{journal}{JCP} \bibinfo{volume}{90},
  \bibinfo{pages}{1007--1023}.
\newblock \DOIprefix\doi{10.1063/1.456153}.
%Type = Article
\bibitem[{{Duvernay} et~al.(2010){Duvernay}, {Dufauret}, {Danger},
  {Theul{\'e}}, {Borget} and {Chiavassa}}]{Duv10}
\bibinfo{author}{{Duvernay}, F.}, \bibinfo{author}{{Dufauret}, V.},
  \bibinfo{author}{{Danger}, G.}, \bibinfo{author}{{Theul{\'e}}, P.},
  \bibinfo{author}{{Borget}, F.}, \bibinfo{author}{{Chiavassa}, T.},
  \bibinfo{year}{2010}.
\newblock \bibinfo{title}{{Chiral molecule formation in interstellar ice
  analogs: alpha-aminoethanol NH$_{2}$CH(CH$_{3}$)OH}}.
\newblock \bibinfo{journal}{AAP} \bibinfo{volume}{523}, \bibinfo{pages}{A79}.
\newblock \DOIprefix\doi{10.1051/0004-6361/201015342}.
%Type = Article
\bibitem[{{Dzenis} et~al.(2022){Dzenis}, {Faure}, {McGuire}, {Remijan},
  {Dagdigian}, {Rist}, {Dawes}, {Quintas-S{\'a}nchez}, {Lique} and
  {Hochlaf}}]{Dze22}
\bibinfo{author}{{Dzenis}, K.}, \bibinfo{author}{{Faure}, A.},
  \bibinfo{author}{{McGuire}, B.A.}, \bibinfo{author}{{Remijan}, A.J.},
  \bibinfo{author}{{Dagdigian}, P.J.}, \bibinfo{author}{{Rist}, C.},
  \bibinfo{author}{{Dawes}, R.}, \bibinfo{author}{{Quintas-S{\'a}nchez}, E.},
  \bibinfo{author}{{Lique}, F.}, \bibinfo{author}{{Hochlaf}, M.},
  \bibinfo{year}{2022}.
\newblock \bibinfo{title}{{Collisional Excitation and Non-LTE Modeling of
  Interstellar Chiral Propylene Oxide}}.
\newblock \bibinfo{journal}{ApJ} \bibinfo{volume}{926}, \bibinfo{pages}{3}.
\newblock \DOIprefix\doi{10.3847/1538-4357/ac43b5},
  \href{http://arxiv.org/abs/2112.08924}{{\tt arXiv:2112.08924}}.
%Type = Article
\bibitem[{{Ellinger} et~al.(2020){Ellinger}, {Pauzat}, {Markovits}, {Allaire}
  and {Guillemin}}]{Ell20}
\bibinfo{author}{{Ellinger}, Y.}, \bibinfo{author}{{Pauzat}, F.},
  \bibinfo{author}{{Markovits}, A.}, \bibinfo{author}{{Allaire}, A.},
  \bibinfo{author}{{Guillemin}, J.C.}, \bibinfo{year}{2020}.
\newblock \bibinfo{title}{{The quest of chirality in the interstellar medium.
  I. Lessons of propylene oxide detection}}.
\newblock \bibinfo{journal}{AAP} \bibinfo{volume}{633}, \bibinfo{pages}{A49}.
\newblock \DOIprefix\doi{10.1051/0004-6361/201936901}.
%Type = Article
\bibitem[{{Endres} et~al.(2016){Endres}, {Schlemmer}, {Schilke}, {Stutzki} and
  {M{\"u}ller}}]{CDMS}
\bibinfo{author}{{Endres}, C.P.}, \bibinfo{author}{{Schlemmer}, S.},
  \bibinfo{author}{{Schilke}, P.}, \bibinfo{author}{{Stutzki}, J.},
  \bibinfo{author}{{M{\"u}ller}, H.S.P.}, \bibinfo{year}{2016}.
\newblock \bibinfo{title}{{The Cologne Database for Molecular Spectroscopy,
  CDMS, in the Virtual Atomic and Molecular Data Centre, VAMDC}}.
\newblock \bibinfo{journal}{Journal of Molecular Spectroscopy}
  \bibinfo{volume}{327}, \bibinfo{pages}{95--104}.
\newblock \DOIprefix\doi{10.1016/j.jms.2016.03.005},
  \href{http://arxiv.org/abs/1603.03264}{{\tt arXiv:1603.03264}}.
%Type = Misc
\bibitem[{Frisch et~al.(2016)Frisch, Trucks, Schlegel, Scuseria, Robb,
  Cheeseman, Scalmani, Barone, Petersson, Nakatsuji, Li, Caricato, Marenich,
  Bloino, Janesko, Gomperts, Mennucci, Hratchian, Ortiz, Izmaylov, Sonnenberg,
  Williams-Young, Ding, Lipparini, Egidi, Goings, Peng, Petrone, Henderson,
  Ranasinghe, Zakrzewski, Gao, Rega, Zheng, Liang, Hada, Ehara, Toyota, Fukuda,
  Hasegawa, Ishida, Nakajima, Honda, Kitao, Nakai, Vreven, Throssell,
  Montgomery, Peralta, Ogliaro, Bearpark, Heyd, Brothers, Kudin, Staroverov,
  Keith, Kobayashi, Normand, Raghavachari, Rendell, Burant, Iyengar, Tomasi,
  Cossi, Millam, Klene, Adamo, Cammi, Ochterski, Martin, Morokuma, Farkas,
  Foresman and Fox}]{G16}
\bibinfo{author}{Frisch, M.J.}, \bibinfo{author}{Trucks, G.W.},
  \bibinfo{author}{Schlegel, H.B.}, \bibinfo{author}{Scuseria, G.E.},
  \bibinfo{author}{Robb, M.A.}, \bibinfo{author}{Cheeseman, J.R.},
  \bibinfo{author}{Scalmani, G.}, \bibinfo{author}{Barone, V.},
  \bibinfo{author}{Petersson, G.A.}, \bibinfo{author}{Nakatsuji, H.},
  \bibinfo{author}{Li, X.}, \bibinfo{author}{Caricato, M.},
  \bibinfo{author}{Marenich, A.V.}, \bibinfo{author}{Bloino, J.},
  \bibinfo{author}{Janesko, B.G.}, \bibinfo{author}{Gomperts, R.},
  \bibinfo{author}{Mennucci, B.}, \bibinfo{author}{Hratchian, H.P.},
  \bibinfo{author}{Ortiz, J.V.}, \bibinfo{author}{Izmaylov, A.F.},
  \bibinfo{author}{Sonnenberg, J.L.}, \bibinfo{author}{Williams-Young, D.},
  \bibinfo{author}{Ding, F.}, \bibinfo{author}{Lipparini, F.},
  \bibinfo{author}{Egidi, F.}, \bibinfo{author}{Goings, J.},
  \bibinfo{author}{Peng, B.}, \bibinfo{author}{Petrone, A.},
  \bibinfo{author}{Henderson, T.}, \bibinfo{author}{Ranasinghe, D.},
  \bibinfo{author}{Zakrzewski, V.G.}, \bibinfo{author}{Gao, J.},
  \bibinfo{author}{Rega, N.}, \bibinfo{author}{Zheng, G.},
  \bibinfo{author}{Liang, W.}, \bibinfo{author}{Hada, M.},
  \bibinfo{author}{Ehara, M.}, \bibinfo{author}{Toyota, K.},
  \bibinfo{author}{Fukuda, R.}, \bibinfo{author}{Hasegawa, J.},
  \bibinfo{author}{Ishida, M.}, \bibinfo{author}{Nakajima, T.},
  \bibinfo{author}{Honda, Y.}, \bibinfo{author}{Kitao, O.},
  \bibinfo{author}{Nakai, H.}, \bibinfo{author}{Vreven, T.},
  \bibinfo{author}{Throssell, K.}, \bibinfo{author}{Montgomery, {Jr.}, J.A.},
  \bibinfo{author}{Peralta, J.E.}, \bibinfo{author}{Ogliaro, F.},
  \bibinfo{author}{Bearpark, M.J.}, \bibinfo{author}{Heyd, J.J.},
  \bibinfo{author}{Brothers, E.N.}, \bibinfo{author}{Kudin, K.N.},
  \bibinfo{author}{Staroverov, V.N.}, \bibinfo{author}{Keith, T.A.},
  \bibinfo{author}{Kobayashi, R.}, \bibinfo{author}{Normand, J.},
  \bibinfo{author}{Raghavachari, K.}, \bibinfo{author}{Rendell, A.P.},
  \bibinfo{author}{Burant, J.C.}, \bibinfo{author}{Iyengar, S.S.},
  \bibinfo{author}{Tomasi, J.}, \bibinfo{author}{Cossi, M.},
  \bibinfo{author}{Millam, J.M.}, \bibinfo{author}{Klene, M.},
  \bibinfo{author}{Adamo, C.}, \bibinfo{author}{Cammi, R.},
  \bibinfo{author}{Ochterski, J.W.}, \bibinfo{author}{Martin, R.L.},
  \bibinfo{author}{Morokuma, K.}, \bibinfo{author}{Farkas, O.},
  \bibinfo{author}{Foresman, J.B.}, \bibinfo{author}{Fox, D.J.},
  \bibinfo{year}{2016}.
\newblock \bibinfo{title}{Gaussian˜16 {R}evision {C}.01}.
\newblock \bibinfo{note}{Gaussian Inc. Wallingford CT}.
%Type = Inproceedings
\bibitem[{{Globus}(2021)}]{Glo21}
\bibinfo{author}{{Globus}, N.}, \bibinfo{year}{2021}.
\newblock \bibinfo{title}{{Homochirality, Cosmic Rays, and Venus}}, in:
  \bibinfo{booktitle}{Venera-D: Venus Cloud Habitability System Workshop}, p.
  \bibinfo{pages}{4057}.
%Type = Article
\bibitem[{{Green} and {Jain}(2010)}]{GJ10}
\bibinfo{author}{{Green}, M.M.}, \bibinfo{author}{{Jain}, V.},
  \bibinfo{year}{2010}.
\newblock \bibinfo{title}{{Homochirality in Life: Two Equal Runners, One
  Tripped}}.
\newblock \bibinfo{journal}{Origins of Life and Evolution of the Biosphere}
  \bibinfo{volume}{40}, \bibinfo{pages}{111--118}.
\newblock \DOIprefix\doi{10.1007/s11084-009-9180-7}.
%Type = Article
\bibitem[{{Grimme} et~al.(2006){Grimme}, {M{\"u}ck-Lichtenfeld},
  {W{\"u}rthwein}, {Ehlers}, {Goumans} and {Lammertsma}}]{B2PLYP}
\bibinfo{author}{{Grimme}, S.}, \bibinfo{author}{{M{\"u}ck-Lichtenfeld}, C.},
  \bibinfo{author}{{W{\"u}rthwein}, E.U.}, \bibinfo{author}{{Ehlers}, A.W.},
  \bibinfo{author}{{Goumans}, T.P.M.}, \bibinfo{author}{{Lammertsma}, K.},
  \bibinfo{year}{2006}.
\newblock \bibinfo{title}{{Consistent Theoretical Description of 1,3-Dipolar
  Cycloaddition Reactions}}.
\newblock \bibinfo{journal}{Journal of Physical Chemistry A}
  \bibinfo{volume}{110}, \bibinfo{pages}{2583--2586}.
\newblock \DOIprefix\doi{10.1021/jp057329x}.
%Type = Misc
\bibitem[{{Haider}(2023)}]{MolDB6}
\bibinfo{author}{{Haider}, N.}, \bibinfo{year}{2023}.
\newblock \bibinfo{title}{Creating a web-based, searchable molecular structure
  database using free software}.
\newblock \URLprefix
  \url{https://homepage.univie.ac.at/norbert.haider/cheminf/moldb.html}.
%Type = Inproceedings
\bibitem[{{Hoare} et~al.(2015){Hoare}, {Perez}, {Bourke}, {Testi},
  {Jimenez-Serra}, {Zarka}, {Siemion}, {van-Langevelde}, {Loinard}, {Anglada},
  {Belloche}, {Bergman}, {Booth}, {Caselli}, {Chandler}, {Codella}, {Hallinan},
  {Lazio}, {Morrison}, {Podio}, {Remijan} and {Tarter}}]{SKACOL}
\bibinfo{author}{{Hoare}, M.}, \bibinfo{author}{{Perez}, L.},
  \bibinfo{author}{{Bourke}, T.L.}, \bibinfo{author}{{Testi}, L.},
  \bibinfo{author}{{Jimenez-Serra}, I.}, \bibinfo{author}{{Zarka}, P.},
  \bibinfo{author}{{Siemion}, A.P.V.}, \bibinfo{author}{{van-Langevelde},
  H.J.}, \bibinfo{author}{{Loinard}, L.}, \bibinfo{author}{{Anglada}, G.},
  \bibinfo{author}{{Belloche}, A.}, \bibinfo{author}{{Bergman}, P.},
  \bibinfo{author}{{Booth}, R.}, \bibinfo{author}{{Caselli}, P.},
  \bibinfo{author}{{Chandler}, C.J.}, \bibinfo{author}{{Codella}, C.},
  \bibinfo{author}{{Hallinan}, G.}, \bibinfo{author}{{Lazio}, J.},
  \bibinfo{author}{{Morrison}, I.S.}, \bibinfo{author}{{Podio}, L.},
  \bibinfo{author}{{Remijan}, A.}, \bibinfo{author}{{Tarter}, J.},
  \bibinfo{year}{2015}.
\newblock \bibinfo{title}{{SKA and the Cradle of Life}}, in:
  \bibinfo{booktitle}{Advancing Astrophysics with the Square Kilometre Array
  (AASKA14)}, p. \bibinfo{pages}{115}.
\newblock \DOIprefix\doi{10.22323/1.215.0115}.
%Type = Article
\bibitem[{{Huang} et~al.(2021){Huang}, {Schwenke} and {Lee}}]{Lee2}
\bibinfo{author}{{Huang}, X.}, \bibinfo{author}{{Schwenke}, D.},
  \bibinfo{author}{{Lee}, T.J.}, \bibinfo{year}{2021}.
\newblock \bibinfo{title}{{What it takes to compute highly accurate
  rovibrational line lists for use in astrochemistry}}.
\newblock \bibinfo{journal}{Accounts of Chemical Research}
  \bibinfo{volume}{54}, \bibinfo{pages}{1311--1321}.
\newblock \DOIprefix\doi{10.15278/isms.2021.TK07}.
%Type = Article
\bibitem[{{Jansen} and {Hess}(1989)}]{Jansen}
\bibinfo{author}{{Jansen}, G.}, \bibinfo{author}{{Hess}, B.A.},
  \bibinfo{year}{1989}.
\newblock \bibinfo{title}{{Relativistic all-electron configuration interaction
  calculations on the gold atom}}.
\newblock \bibinfo{journal}{Chemical Physics Letters} \bibinfo{volume}{160},
  \bibinfo{pages}{507--513}.
\newblock \DOIprefix\doi{10.1016/0009-2614(89)80054-5}.
%Type = Misc
\bibitem[{{Johnson III}(2022)}]{nist}
\bibinfo{author}{{Johnson III}, R.}, \bibinfo{year}{2022}.
\newblock \bibinfo{title}{Nist standard reference database number 101}.
\newblock \bibinfo{note}{Available at \url{http:// https://cccbdb.nist.gov/}}.
%Type = Article
\bibitem[{{Kendall} et~al.(1992){Kendall}, {Dunning} and {Harrison}}]{Kendall}
\bibinfo{author}{{Kendall}, R.A.}, \bibinfo{author}{{Dunning}, Thom~H., J.},
  \bibinfo{author}{{Harrison}, R.J.}, \bibinfo{year}{1992}.
\newblock \bibinfo{title}{{Electron affinities of the first-row atoms
  revisited. Systematic basis sets and wave functions}}.
\newblock \bibinfo{journal}{JCP} \bibinfo{volume}{96},
  \bibinfo{pages}{6796--6806}.
\newblock \DOIprefix\doi{10.1063/1.462569}.
%Type = Article
\bibitem[{{Lankhaar}(2022)}]{Lankhaar_2022}
\bibinfo{author}{{Lankhaar}, B.}, \bibinfo{year}{2022}.
\newblock \bibinfo{title}{{Detecting chiral asymmetry in the interstellar
  medium using propylene oxide}}.
\newblock \bibinfo{journal}{AAP} \bibinfo{volume}{666}, \bibinfo{pages}{A126}.
\newblock \DOIprefix\doi{10.1051/0004-6361/202244295},
  \href{http://arxiv.org/abs/2207.02888}{{\tt arXiv:2207.02888}}.
%Type = Article
\bibitem[{{Lee} and {McCarthy}(2020)}]{Lee2020}
\bibinfo{author}{{Lee}, K.L.K.}, \bibinfo{author}{{McCarthy}, M.},
  \bibinfo{year}{2020}.
\newblock \bibinfo{title}{{Bayesian Analysis of Theoretical Rot ational
  Constants from Low-Cost Electronic Structure Methods}}.
\newblock \bibinfo{journal}{Journal of Physical Chemistry A}
  \bibinfo{volume}{124}, \bibinfo{pages}{898--910}.
\newblock \DOIprefix\doi{10.1021/acs.jpca.9b09982}.
%Type = Article
\bibitem[{{Martin} and {Lee}(1996)}]{ML96}
\bibinfo{author}{{Martin}, J.M.L.}, \bibinfo{author}{{Lee}, T.J.},
  \bibinfo{year}{1996}.
\newblock \bibinfo{title}{{The atomization energy and proton affinity of NH
  $_{3}$. An ab initio calibration study}}.
\newblock \bibinfo{journal}{Chemical Physics Letters} \bibinfo{volume}{258},
  \bibinfo{pages}{136--143}.
\newblock \DOIprefix\doi{10.1016/0009-2614(96)00658-6}.
%Type = Article
\bibitem[{{Martin} et~al.(1994){Martin}, {Taylor}, {Fran{\c{c}}ois} and
  {Gijbels}}]{MT94}
\bibinfo{author}{{Martin}, J.M.L.}, \bibinfo{author}{{Taylor}, P.R.},
  \bibinfo{author}{{Fran{\c{c}}ois}, J.P.}, \bibinfo{author}{{Gijbels}, R.},
  \bibinfo{year}{1994}.
\newblock \bibinfo{title}{{Ab initio study of the spectroscopy and
  thermochemistry of the C $_{2}$N and CN $_{2}$ molecules}}.
\newblock \bibinfo{journal}{Chemical Physics Letters} \bibinfo{volume}{226},
  \bibinfo{pages}{475--483}.
\newblock \DOIprefix\doi{10.1016/0009-2614(94)00758-6}.
%Type = Article
\bibitem[{{McGuire}(2022)}]{McG22}
\bibinfo{author}{{McGuire}, B.A.}, \bibinfo{year}{2022}.
\newblock \bibinfo{title}{{2021 Census of Interstellar, Circumstellar,
  Extragalactic, Protoplanetary Disk, and Exoplanetary Molecules}}.
\newblock \bibinfo{journal}{ApJs} \bibinfo{volume}{259}, \bibinfo{pages}{30}.
\newblock \DOIprefix\doi{10.3847/1538-4365/ac2a48},
  \href{http://arxiv.org/abs/2109.13848}{{\tt arXiv:2109.13848}}.
%Type = Article
\bibitem[{{McGuire} et~al.(2020){McGuire}, {Burkhardt}, {Loomis},
  {Shingledecker}, {Kelvin Lee}, {Charnley}, {Cordiner}, {Herbst}, {Kalenskii},
  {Momjian}, {Willis}, {Xue}, {Remijan} and {McCarthy}}]{GOTHAM}
\bibinfo{author}{{McGuire}, B.A.}, \bibinfo{author}{{Burkhardt}, A.M.},
  \bibinfo{author}{{Loomis}, R.A.}, \bibinfo{author}{{Shingledecker}, C.N.},
  \bibinfo{author}{{Kelvin Lee}, K.L.}, \bibinfo{author}{{Charnley}, S.B.},
  \bibinfo{author}{{Cordiner}, M.A.}, \bibinfo{author}{{Herbst}, E.},
  \bibinfo{author}{{Kalenskii}, S.}, \bibinfo{author}{{Momjian}, E.},
  \bibinfo{author}{{Willis}, E.R.}, \bibinfo{author}{{Xue}, C.},
  \bibinfo{author}{{Remijan}, A.J.}, \bibinfo{author}{{McCarthy}, M.C.},
  \bibinfo{year}{2020}.
\newblock \bibinfo{title}{{Early Science from GOTHAM: Project Overview,
  Methods, and the Detection of Interstellar Propargyl Cyanide (HCCCH$_{2}$CN)
  in TMC-1}}.
\newblock \bibinfo{journal}{ApJl} \bibinfo{volume}{900}, \bibinfo{pages}{L10}.
\newblock \DOIprefix\doi{10.3847/2041-8213/aba632},
  \href{http://arxiv.org/abs/2008.12349}{{\tt arXiv:2008.12349}}.
%Type = Inproceedings
\bibitem[{{McGuire} et~al.(2018){McGuire}, {Carroll} and {Garrod}}]{McG18}
\bibinfo{author}{{McGuire}, B.A.}, \bibinfo{author}{{Carroll}, P.B.},
  \bibinfo{author}{{Garrod}, R.T.}, \bibinfo{year}{2018}.
\newblock \bibinfo{title}{{Prebiotic Molecules}}, in:
  \bibinfo{editor}{{Murphy}, E.} (Ed.), \bibinfo{booktitle}{Science with a Next
  Generation Very Large Array}, p. \bibinfo{pages}{245}.
%Type = Article
\bibitem[{{McGuire} et~al.(2016){McGuire}, {Carroll}, {Loomis}, {Finneran},
  {Jewell}, {Remijan} and {Blake}}]{McG16}
\bibinfo{author}{{McGuire}, B.A.}, \bibinfo{author}{{Carroll}, P.B.},
  \bibinfo{author}{{Loomis}, R.A.}, \bibinfo{author}{{Finneran}, I.A.},
  \bibinfo{author}{{Jewell}, P.R.}, \bibinfo{author}{{Remijan}, A.J.},
  \bibinfo{author}{{Blake}, G.A.}, \bibinfo{year}{2016}.
\newblock \bibinfo{title}{{Discovery of the interstellar chiral molecule
  propylene oxide (CH$_{3}$CHCH$_{2}$O)}}.
\newblock \bibinfo{journal}{Science} \bibinfo{volume}{352},
  \bibinfo{pages}{1449--1452}.
\newblock \DOIprefix\doi{10.1126/science.aae0328},
  \href{http://arxiv.org/abs/1606.07483}{{\tt arXiv:1606.07483}}.
%Type = Article
\bibitem[{{Mesko} et~al.(2017){Mesko}, {Zou}, {Carroll} and {Weaver}}]{Mesko}
\bibinfo{author}{{Mesko}, A.}, \bibinfo{author}{{Zou}, L.},
  \bibinfo{author}{{Carroll}, P.}, \bibinfo{author}{{Weaver}, S.},
  \bibinfo{year}{2017}.
\newblock \bibinfo{title}{{Millimeter and submillimeter spectrum of propylene
  oxide}}.
\newblock \bibinfo{journal}{Journal of Molecular Spectroscopy}
  \bibinfo{volume}{335}, \bibinfo{pages}{49--53}.
%Type = Article
\bibitem[{{M{\o}llendal} et~al.(2012){M{\o}llendal}, {Margul{\`e}s},
  {Belloche}, {Motiyenko}, {Konovalov}, {Menten} and {Guillemin}}]{Mol12}
\bibinfo{author}{{M{\o}llendal}, H.}, \bibinfo{author}{{Margul{\`e}s}, L.},
  \bibinfo{author}{{Belloche}, A.}, \bibinfo{author}{{Motiyenko}, R.A.},
  \bibinfo{author}{{Konovalov}, A.}, \bibinfo{author}{{Menten}, K.M.},
  \bibinfo{author}{{Guillemin}, J.C.}, \bibinfo{year}{2012}.
\newblock \bibinfo{title}{{Rotational spectrum of a chiral amino acid
  precursor, 2-aminopropionitrile, and searches for it in Sagittarius B2(N)}}.
\newblock \bibinfo{journal}{AAP} \bibinfo{volume}{538}, \bibinfo{pages}{A51}.
\newblock \DOIprefix\doi{10.1051/0004-6361/201116838}.
%Type = Article
\bibitem[{{Morrison} et~al.(2023){Morrison}, {Crosse}, {Sleap}, {Wayth},
  {Williams}, {Johnston-Hollitt}, {Jones}, {Tingay}, {Walker} and
  {Williams}}]{Mor23}
\bibinfo{author}{{Morrison}, I.S.}, \bibinfo{author}{{Crosse}, B.},
  \bibinfo{author}{{Sleap}, G.}, \bibinfo{author}{{Wayth}, R.B.},
  \bibinfo{author}{{Williams}, A.}, \bibinfo{author}{{Johnston-Hollitt}, M.},
  \bibinfo{author}{{Jones}, J.}, \bibinfo{author}{{Tingay}, S.J.},
  \bibinfo{author}{{Walker}, M.}, \bibinfo{author}{{Williams}, L.},
  \bibinfo{year}{2023}.
\newblock \bibinfo{title}{{MWAX: A new correlator for the Murchison Widefield
  Array}}.
\newblock \bibinfo{journal}{PASA} \bibinfo{volume}{40}, \bibinfo{pages}{e019}.
\newblock \DOIprefix\doi{10.1017/pasa.2023.15},
  \href{http://arxiv.org/abs/2303.11557}{{\tt arXiv:2303.11557}}.
%Type = Article
\bibitem[{{M{\"u}ller} and {Bauder}(1996)}]{MB96}
\bibinfo{author}{{M{\"u}ller}, F.}, \bibinfo{author}{{Bauder}, A.},
  \bibinfo{year}{1996}.
\newblock \bibinfo{title}{{Microwave Spectrum, Quadrupole Coupling Constants,
  and Dipole Moment of Oxiranecarbonitrile}}.
\newblock \bibinfo{journal}{Journal of Molecular Spectroscopy}
  \bibinfo{volume}{179}, \bibinfo{pages}{61--64}.
\newblock \DOIprefix\doi{10.1006/jmsp.1996.0183}.
%Type = Article
\bibitem[{{Myrgorodska} et~al.(2015){Myrgorodska}, {Meinert}, {Martins}, {Le
  Sergeant d'Hendecourt} and {Meierhenrich}}]{Myr15}
\bibinfo{author}{{Myrgorodska}, I.}, \bibinfo{author}{{Meinert}, C.},
  \bibinfo{author}{{Martins}, Z.}, \bibinfo{author}{{Le Sergeant d'Hendecourt},
  L.}, \bibinfo{author}{{Meierhenrich}, U.J.}, \bibinfo{year}{2015}.
\newblock \bibinfo{title}{{Molek{\"u}lchiralit{\"a}t in Meteoriten und
  interstellarem Eis und das Chiralit{\"a}tsexperiment an Bord der
  Kometenmission Rosetta der ESA}}.
\newblock \bibinfo{journal}{Angewandte Chemie} \bibinfo{volume}{127},
  \bibinfo{pages}{1420--1430}.
\newblock \DOIprefix\doi{10.1002/ange.201409354}.
%Type = Article
\bibitem[{{Pasteur}(1848)}]{Pasteur}
\bibinfo{author}{{Pasteur}, L.}, \bibinfo{year}{1848}.
\newblock \bibinfo{title}{{Ueber die Krystallisation des Schwefels}}.
\newblock \bibinfo{journal}{Annalen der Physik} \bibinfo{volume}{150},
  \bibinfo{pages}{94}.
\newblock \DOIprefix\doi{10.1002/andp.18491500510}.
%Type = Article
\bibitem[{{Paulive} et~al.(2021){Paulive}, {Shingledecker} and
  {Herbst}}]{Paulive2021}
\bibinfo{author}{{Paulive}, A.}, \bibinfo{author}{{Shingledecker}, C.N.},
  \bibinfo{author}{{Herbst}, E.}, \bibinfo{year}{2021}.
\newblock \bibinfo{title}{{The role of radiolysis in the modelling of
  C$_{2}$H$_{4}$O$_{2}$ isomers and dimethyl ether in cold dark clouds}}.
\newblock \bibinfo{journal}{MNRAS} \bibinfo{volume}{500},
  \bibinfo{pages}{3414--3424}.
\newblock \DOIprefix\doi{10.1093/mnras/staa3458},
  \href{http://arxiv.org/abs/2011.02023}{{\tt arXiv:2011.02023}}.
%Type = Article
\bibitem[{{Perley} et~al.(2011){Perley}, {Chandler}, {Butler} and
  {Wrobel}}]{evla_11}
\bibinfo{author}{{Perley}, R.A.}, \bibinfo{author}{{Chandler}, C.J.},
  \bibinfo{author}{{Butler}, B.J.}, \bibinfo{author}{{Wrobel}, J.M.},
  \bibinfo{year}{2011}.
\newblock \bibinfo{title}{{The Expanded Very Large Array: A New Telescope for
  New Science}}.
\newblock \bibinfo{journal}{ApJl} \bibinfo{volume}{739}, \bibinfo{pages}{L1}.
\newblock \DOIprefix\doi{10.1088/2041-8205/739/1/L1},
  \href{http://arxiv.org/abs/1106.0532}{{\tt arXiv:1106.0532}}.
%Type = Article
\bibitem[{{Purvis} and {Bartlett}(1982)}]{CCSD1}
\bibinfo{author}{{Purvis}, George~D., I.}, \bibinfo{author}{{Bartlett}, R.J.},
  \bibinfo{year}{1982}.
\newblock \bibinfo{title}{{A full coupled-cluster singles and doubles model:
  The inclusion of disconnected triples}}.
\newblock \bibinfo{journal}{JCP} \bibinfo{volume}{76},
  \bibinfo{pages}{1910--1918}.
\newblock \DOIprefix\doi{10.1063/1.443164}.
%Type = Article
\bibitem[{Puzzarini(2017)}]{CP17}
\bibinfo{author}{Puzzarini, C.}, \bibinfo{year}{2017}.
\newblock \bibinfo{title}{Astronomical complex organic molecules: Quantum
  chemistry meets rotational spectroscopy}.
\newblock \bibinfo{journal}{International Journal of Quantum Chemistry}
  \bibinfo{volume}{117}, \bibinfo{pages}{129--138}.
\newblock \URLprefix
  \url{https://onlinelibrary.wiley.com/doi/abs/10.1002/qua.25284},
  \DOIprefix\doi{https://doi.org/10.1002/qua.25284},
  \href{http://arxiv.org/abs/https://onlinelibrary.wiley.com/doi/pdf/10.1002/qua.25284}{{\tt
  arXiv:https://onlinelibrary.wiley.com/doi/pdf/10.1002/qua.25284}}.
%Type = Article
\bibitem[{{Puzzarini} and {Stanton}(2023)}]{CS23}
\bibinfo{author}{{Puzzarini}, C.}, \bibinfo{author}{{Stanton}, J.F.},
  \bibinfo{year}{2023}.
\newblock \bibinfo{title}{{Connections between the accuracy of rotational
  constants and equilibrium molecular structures}}.
\newblock \bibinfo{journal}{Physical Chemistry Chemical Physics (Incorporating
  Faraday Transactions)} \bibinfo{volume}{25}, \bibinfo{pages}{1421--1429}.
\newblock \DOIprefix\doi{10.1039/D2CP04706C}.
%Type = Article
\bibitem[{{Raghavachari} et~al.(1989){Raghavachari}, {Trucks}, {Pople} and
  {Head-Gordon}}]{CCSD2}
\bibinfo{author}{{Raghavachari}, K.}, \bibinfo{author}{{Trucks}, G.W.},
  \bibinfo{author}{{Pople}, J.A.}, \bibinfo{author}{{Head-Gordon}, M.},
  \bibinfo{year}{1989}.
\newblock \bibinfo{title}{{A fifth-order perturbation comparison of electron
  correlation theories}}.
\newblock \bibinfo{journal}{Chemical Physics Letters} \bibinfo{volume}{157},
  \bibinfo{pages}{479--483}.
\newblock \DOIprefix\doi{10.1016/S0009-2614(89)87395-6}.
%Type = Article
\bibitem[{{Rastoltseva} et~al.(2010){Rastoltseva}, {Bataev} and
  {Godunov}}]{RBG10}
\bibinfo{author}{{Rastoltseva}, E.V.}, \bibinfo{author}{{Bataev}, V.A.},
  \bibinfo{author}{{Godunov}, I.A.}, \bibinfo{year}{2010}.
\newblock \bibinfo{title}{{Structure and conformational dynamics of
  oxiranecarboxaldehyde in the ground and excited electronic states}}.
\newblock \bibinfo{journal}{Journal of Molecular Structure}
  \bibinfo{volume}{978}, \bibinfo{pages}{269--278}.
\newblock \DOIprefix\doi{10.1016/j.molstruc.2010.03.007}.
%Type = Inproceedings
\bibitem[{{Remijan} et~al.(2007){Remijan}, {Markwick-Kemper} and {ALMA Working
  Group on Spectral Line Frequencies}}]{SPLAT}
\bibinfo{author}{{Remijan}, A.J.}, \bibinfo{author}{{Markwick-Kemper}, A.},
  \bibinfo{author}{{ALMA Working Group on Spectral Line Frequencies}},
  \bibinfo{year}{2007}.
\newblock \bibinfo{title}{{Splatalogue: Database for Astronomical
  Spectroscopy}}, in: \bibinfo{booktitle}{American Astronomical Society Meeting
  Abstracts}, p. \bibinfo{pages}{132.11}.
%Type = Article
\bibitem[{{Sch{\"o}ier} et~al.(2005){Sch{\"o}ier}, {van der Tak}, {van
  Dishoeck} and {Black}}]{Sho05}
\bibinfo{author}{{Sch{\"o}ier}, F.L.}, \bibinfo{author}{{van der Tak}, F.F.S.},
  \bibinfo{author}{{van Dishoeck}, E.F.}, \bibinfo{author}{{Black}, J.H.},
  \bibinfo{year}{2005}.
\newblock \bibinfo{title}{{An atomic and molecular database for analysis of
  submillimetre line observations}}.
\newblock \bibinfo{journal}{AAP} \bibinfo{volume}{432},
  \bibinfo{pages}{369--379}.
\newblock \DOIprefix\doi{10.1051/0004-6361:20041729},
  \href{http://arxiv.org/abs/astro-ph/0411110}{{\tt arXiv:astro-ph/0411110}}.
%Type = Article
\bibitem[{{Simmie}(2022)}]{Simmie22}
\bibinfo{author}{{Simmie}, J.M.}, \bibinfo{year}{2022}.
\newblock \bibinfo{title}{{C2H5NO Isomers: From Acetamide to 1,2-Oxazetidine
  and Beyond}}.
\newblock \bibinfo{journal}{Journal of Physical Chemistry A}
  \bibinfo{volume}{126}, \bibinfo{pages}{924--939}.
\newblock \DOIprefix\doi{10.1021/acs.jpca.1c09984}.
%Type = Article
\bibitem[{{Tennyson} and {Yurchenko}(2012)}]{ExoMOL}
\bibinfo{author}{{Tennyson}, J.}, \bibinfo{author}{{Yurchenko}, S.N.},
  \bibinfo{year}{2012}.
\newblock \bibinfo{title}{{ExoMol: molecular line lists for exoplanet and other
  atmospheres}}.
\newblock \bibinfo{journal}{MNRAS} \bibinfo{volume}{425},
  \bibinfo{pages}{21--33}.
\newblock \DOIprefix\doi{10.1111/j.1365-2966.2012.21440.x},
  \href{http://arxiv.org/abs/1204.0124}{{\tt arXiv:1204.0124}}.
%Type = Phdthesis
\bibitem[{{Tremblay}(2018)}]{Tre18}
\bibinfo{author}{{Tremblay}, C.}, \bibinfo{year}{2018}.
\newblock \bibinfo{title}{{A Search for Molecules at Low Frequency with the
  Murchison Widefield Array}}.
\newblock Ph.D. thesis. Curtin University, Australia.
%Type = Article
\bibitem[{{Tremblay} et~al.(2020){Tremblay}, {Gray}, {Hurley-Walker}, {Green},
  {Dawson}, {Dickey}, {Jones}, {Tingay} and {Wong}}]{Tre20}
\bibinfo{author}{{Tremblay}, C.D.}, \bibinfo{author}{{Gray}, M.D.},
  \bibinfo{author}{{Hurley-Walker}, N.}, \bibinfo{author}{{Green}, J.A.},
  \bibinfo{author}{{Dawson}, J.R.}, \bibinfo{author}{{Dickey}, J.M.},
  \bibinfo{author}{{Jones}, P.A.}, \bibinfo{author}{{Tingay}, S.J.},
  \bibinfo{author}{{Wong}, O.I.}, \bibinfo{year}{2020}.
\newblock \bibinfo{title}{{Nitric Oxide and Other Molecules: Molecular Modeling
  and Low-frequency Exploration Using the Murchison Widefield Array}}.
\newblock \bibinfo{journal}{ApJ} \bibinfo{volume}{905}, \bibinfo{pages}{65}.
\newblock \DOIprefix\doi{10.3847/1538-4357/abc33a}.
%Type = Article
\bibitem[{{Turner} et~al.(1975){Turner}, {Liszt}, {Kaifu} and
  {Kisliakov}}]{Turner_75}
\bibinfo{author}{{Turner}, B.E.}, \bibinfo{author}{{Liszt}, H.S.},
  \bibinfo{author}{{Kaifu}, N.}, \bibinfo{author}{{Kisliakov}, A.G.},
  \bibinfo{year}{1975}.
\newblock \bibinfo{title}{{Microwave detection of interstellar cyanamide.}}
\newblock \bibinfo{journal}{APJl} \bibinfo{volume}{201},
  \bibinfo{pages}{L149--L152}.
\newblock \DOIprefix\doi{10.1086/181963}.
%Type = Article
\bibitem[{{Wayth} et~al.(2018){Wayth}, {Tingay}, {Trott}, {Emrich},
  {Johnston-Hollitt}, {McKinley}, {Gaensler}, {Beardsley}, {Booler}, {Crosse},
  {Franzen}, {Horsley}, {Kaplan}, {Kenney}, {Morales}, {Pallot}, {Sleap},
  {Steele}, {Walker}, {Williams}, {Wu}, {Cairns}, {Filipovic}, {Johnston},
  {Murphy}, {Quinn}, {Staveley-Smith}, {Webster} and {Wyithe}}]{Way18}
\bibinfo{author}{{Wayth}, R.B.}, \bibinfo{author}{{Tingay}, S.J.},
  \bibinfo{author}{{Trott}, C.M.}, \bibinfo{author}{{Emrich}, D.},
  \bibinfo{author}{{Johnston-Hollitt}, M.}, \bibinfo{author}{{McKinley}, B.},
  \bibinfo{author}{{Gaensler}, B.M.}, \bibinfo{author}{{Beardsley}, A.P.},
  \bibinfo{author}{{Booler}, T.}, \bibinfo{author}{{Crosse}, B.},
  \bibinfo{author}{{Franzen}, T.M.O.}, \bibinfo{author}{{Horsley}, L.},
  \bibinfo{author}{{Kaplan}, D.L.}, \bibinfo{author}{{Kenney}, D.},
  \bibinfo{author}{{Morales}, M.F.}, \bibinfo{author}{{Pallot}, D.},
  \bibinfo{author}{{Sleap}, G.}, \bibinfo{author}{{Steele}, K.},
  \bibinfo{author}{{Walker}, M.}, \bibinfo{author}{{Williams}, A.},
  \bibinfo{author}{{Wu}, C.}, \bibinfo{author}{{Cairns}, I.H.},
  \bibinfo{author}{{Filipovic}, M.D.}, \bibinfo{author}{{Johnston}, S.},
  \bibinfo{author}{{Murphy}, T.}, \bibinfo{author}{{Quinn}, P.},
  \bibinfo{author}{{Staveley-Smith}, L.}, \bibinfo{author}{{Webster}, R.},
  \bibinfo{author}{{Wyithe}, J.S.B.}, \bibinfo{year}{2018}.
\newblock \bibinfo{title}{{The Phase II Murchison Widefield Array: Design
  overview}}.
\newblock \bibinfo{journal}{PASA} \bibinfo{volume}{35}, \bibinfo{pages}{e033}.
\newblock \DOIprefix\doi{10.1017/pasa.2018.37},
  \href{http://arxiv.org/abs/1809.06466}{{\tt arXiv:1809.06466}}.
%Type = Article
\bibitem[{{Werner} et~al.(2020){Werner}, {Knowles}, {Manby}, {Black}, {Doll},
  {He{\ss}elmann}, {Kats}, {K{\"o}hn}, {Korona}, {Kreplin}, {Ma}, {Miller},
  {Mitrushchenkov}, {Peterson}, {Polyak}, {Rauhut} and {Sibaev}}]{molpro}
\bibinfo{author}{{Werner}, H.J.}, \bibinfo{author}{{Knowles}, P.J.},
  \bibinfo{author}{{Manby}, F.R.}, \bibinfo{author}{{Black}, J.A.},
  \bibinfo{author}{{Doll}, K.}, \bibinfo{author}{{He{\ss}elmann}, A.},
  \bibinfo{author}{{Kats}, D.}, \bibinfo{author}{{K{\"o}hn}, A.},
  \bibinfo{author}{{Korona}, T.}, \bibinfo{author}{{Kreplin}, D.A.},
  \bibinfo{author}{{Ma}, Q.}, \bibinfo{author}{{Miller}, Thomas~F., I.},
  \bibinfo{author}{{Mitrushchenkov}, A.}, \bibinfo{author}{{Peterson}, K.A.},
  \bibinfo{author}{{Polyak}, I.}, \bibinfo{author}{{Rauhut}, G.},
  \bibinfo{author}{{Sibaev}, M.}, \bibinfo{year}{2020}.
\newblock \bibinfo{title}{{The Molpro quantum chemistry package}}.
\newblock \bibinfo{journal}{JCP} \bibinfo{volume}{152},
  \bibinfo{pages}{144107}.
\newblock \DOIprefix\doi{10.1063/5.0005081}.
%Type = Article
\bibitem[{{Zapata Trujillo} et~al.(2021){Zapata Trujillo}, {Syme}, {Rowell},
  {Burns}, {Clark}, {Gorman}, {Jacob}, {Kapodistrias}, {Kedziora}, {Lempriere},
  {Medcraft}, {O'Sullivan}, {Robertson}, {Soares}, {Steller}, {Teece},
  {Tremblay}, {Sousa-Silva} and {McKemmish}}]{Zapata_2021}
\bibinfo{author}{{Zapata Trujillo}, J.C.}, \bibinfo{author}{{Syme}, A.M.},
  \bibinfo{author}{{Rowell}, K.N.}, \bibinfo{author}{{Burns}, B.P.},
  \bibinfo{author}{{Clark}, E.S.}, \bibinfo{author}{{Gorman}, M.N.},
  \bibinfo{author}{{Jacob}, L.S.D.}, \bibinfo{author}{{Kapodistrias}, P.},
  \bibinfo{author}{{Kedziora}, D.J.}, \bibinfo{author}{{Lempriere}, F.A.R.},
  \bibinfo{author}{{Medcraft}, C.}, \bibinfo{author}{{O'Sullivan}, J.},
  \bibinfo{author}{{Robertson}, E.G.}, \bibinfo{author}{{Soares}, G.G.},
  \bibinfo{author}{{Steller}, L.}, \bibinfo{author}{{Teece}, B.L.},
  \bibinfo{author}{{Tremblay}, C.D.}, \bibinfo{author}{{Sousa-Silva}, C.},
  \bibinfo{author}{{McKemmish}, L.K.}, \bibinfo{year}{2021}.
\newblock \bibinfo{title}{{Computational Infrared Spectroscopy of 956
  Phosphorus-bearing Molecules}}.
\newblock \bibinfo{journal}{Frontiers in Astronomy and Space Sciences}
  \bibinfo{volume}{8}, \bibinfo{pages}{43}.
\newblock \DOIprefix\doi{10.3389/fspas.2021.639068},
  \href{http://arxiv.org/abs/2105.08897}{{\tt arXiv:2105.08897}}.

\end{thebibliography}



%% For example:

%% \bibitem[Aladro et al.(2015)]{Aladro15} Aladro, R., Martín, S., Riquelme, D., et al. 2015, \aas, 579, A101

%%\end{thebibliography}

\end{document}